\documentclass[aps,prb,twocolumn,showpacs,superscriptaddress,groupedaddress]{revtex4}  
\usepackage{dcolumn}   
\usepackage{bm}        
\usepackage{amssymb}   
\usepackage{amsmath}	 
\usepackage{hyperref}
\hypersetup{colorlinks=true,urlcolor=blue,citecolor=blue,linkcolor=blue,breaklinks=true}

\usepackage{tikz}
\usetikzlibrary{decorations.markings} 
\usetikzlibrary{patterns} 

\usepackage{graphicx}  

\usepackage{amsfonts}

\usepackage[normalem]{ulem}

\usepackage{lipsum} 

\newcommand{\mat}[1]{\underline{\underline{#1}}}
\newcommand{\ov}[1]{\overline{#1}}

\newcommand{\red}[1]{{\color{red}{#1}}}

\begin{document}

\title{Truncated Unity Parquet Solver}

\author{Christian J. Eckhardt}
\email{christian.eckhardt@rwth-aachen.de}
\affiliation{Institute for Solid State Physics, TU Wien, 1040 Vienna, Austria}
\affiliation{Institute for Theoretical Solid State Physics, RWTH Aachen University, 52074 Aachen, Germany}
\author{Carsten Honerkamp}
\affiliation{Institute for Theoretical Solid State Physics, RWTH Aachen University, 52074 Aachen, Germany}
\affiliation{JARA-FIT, Jülich Aachen Research Alliance - Fundamentals of Future Information Technology, Germany}
\author{Karsten Held}
\affiliation{Institute for Solid State Physics, TU Wien, 1040 Vienna, Austria}
\author{Anna Kauch}
\email{kauch@ifp.tuwien.ac.at}
\affiliation{Institute for Solid State Physics, TU Wien, 1040 Vienna, Austria}

\date{\today}

\begin{abstract}
We present an implementation of a truncated unity parquet solver (TUPS) which solves the parquet equations using  a truncated form-factor basis for the fermionic momenta. This way fluctuations from different scattering channels are treated on an equal footing.  The essentially linear scaling of computational costs in the number of untruncated bosonic momenta allows us to treat system sizes of up to $76 \times 76$ discrete lattice momenta, unprecedented by previous unbiased methods that include the frequency dependence of the vertex.
With TUPS, we provide  the first numerical evidence that the parquet approximation might indeed respect the  Mermin-Wagner theorem and further systematically analyze the convergence with respect to the number of form factors. Using a single form factor seems to qualitatively describe the physics of the half-filled Hubbard model correctly, including the pseudogap behaviour.
Quantitatively, using a single or a few form factors only is not sufficient at lower temperatures or stronger coupling.
\end{abstract}

\pacs{}
\maketitle

\section{Introduction}
\label{sec:introduction}

Strongly correlated electron systems often show  very rich phase diagrams where, e.g., magnetic,  charge density wave, and superconducting  phases lie close by. The interplay of such competing orders and fluctuations may give rise to fascinating new physics, possibly also to  high-temperature superconductivity
through e.g.~spin-fluctuation mediated superconductivity
\cite{Scalapino12}.
Hence a proper understanding of correlated systems often necessitates methods where the competition between different instabilities (magnetic, charge or pairing) is treated in an unbiased way, i.e., without assuming that one of these scattering channels dominates.

Beside numerically exact methods such as exact diagonalization (ED) \cite{Nakano2007} and lattice quantum Monte Carlo (QMC) simulations \cite{Blankenbecler1981,White1989} for finite systems, and the density matrix renormalization group (DMRG)\cite{Schollwock2011} for one dimension, several diagrammatic approaches also provide an unbiased framework.
Among these are renormalization group (RG) methods such as functional RG (fRG)~\cite{Metzner12} and parquet equations~\cite{parquetII,Vasiliev98} based methods. The  latter include the parquet approximation (PA)~\cite{Bickers04} and diagrammatic extensions of dynamical mean-field theory (DMFT)~\cite{RMPVertex} such as parquet dynamical vertex approximation (D$\Gamma$A)~\cite{Valli2015,Li2016,Kauch2019}, parquet dual fermion (DF)~\cite{Astretsov2019,Astleithner2019} and a combination with fRG: DMF$^2$RG~\cite{Taranto2014}. A close connection between fRG and parquet equations has been established recently in Refs.~\onlinecite{Kugler18, Kugler18a}. 

 Although the numerical effort of diagrammatic methods scales only polynomially in the system size, the computational effort needed to solve the parquet equations  is still enormous, even in the simple case of the 2D Hubbard model.
To cope with prohibitively huge memory and computing time requirements, two different strategies have been hitherto used: (i) neglecting partially or fully the frequency dependence of the two-particle vertices~\cite{Lichtenstein17,Metzner12,Astretsov2019,Janis2017}; (ii) neglecting the dependence on the fermionic momentum, or employing a rather coarse grid in all three momenta of the vertex~\cite{Tam2013,Yang2009,Li2016,Li2019,Kauch2019,Pudleiner2019a}. 

Neglecting the frequency dependence leads to a strong shift of the phase transitions to higher critical temperatures \cite{Astretsov2019,Tagliavini19} including the superconducting $T_c$ \cite{Kitatani2018,Tagliavini19}. The full  frequency dependence also furthers odd-frequency singlet superconducting fluctuations at larger dopings in the Hubbard model~\cite{Kauch2019}.
On the other hand, compromising the momentum resolution leads to an inadequate description of superconductivity~\cite{Gull2013}, magnetic susceptibility~\cite{Kauch2019}, charge ordering~\cite{Zheng2017, Ponsioen2019}  and the pseudogap~\cite{Kozik2018}.
In order to gain sufficient resolution at least in the bosonic transfer momentum ${\vec q}$, ladder based approximations are used, where the non-local coupling of different channels (particle-hole and particle-particle) is neglected, as e.g.\ in ladder D$\Gamma$A~\cite{Toschi2007,Katanin2009} and ladder DF~\cite{Otsuki2014a}. The computational advantage mainly lies in the fact that there is only  one momentum dependence, which even makes realistic material calculations possible, e.g., with  {\it ab initio} D$\Gamma$A~\cite{Galler2016}.
This way however the unbiased interplay between magnetic, density and pairing fluctuations is lost to a large degree.  


 An approach that in principle does not compromise, neither on the frequency nor the momentum dependence, is 
the truncated unity (TU) approach that has been formally introduced in Refs.~\onlinecite{Lichtenstein17,Eckhardt18}. The TU approach  exploits the fact that the two-particle vertices tend to have a simpler structure in the so-called fermionic momenta than in the transfer bosonic momentum~\cite{Husemann09}.
 The dependence on the fermionic momenta is captured instead by a form-factor expansion, using a basis set that includes increasingly remote lattice sites, or increasingly complicated momentum structures.
Such a form-factor expansion had, to some extent, been used  already previously in fRG~\cite{Husemann09,Metzner12,Maier13,Lichtenstein17,Vilardi19,Tagliavini19, Wang2013}. However, to the best of our knowledge, all 
previous approaches take only one or a few handpicked form factors into account, or neglect the frequency dependence of the vertices~\cite{Lichtenstein17}.

In this paper, we implement a numerical iterative solution of the parquet equations with a truncated unity to an arbitrary number of form factors. 
Beyond Refs.~\onlinecite{Lichtenstein17,Eckhardt18}, we further present an unbiased formulation of the Schwinger-Dyson equation in the form-factor basis with what  we will later call ``channel native'' momenta, and argue for its importance compared to the na\"ive implementation. The method as well as the code that is made available on GitHub~\cite{TUPS} are
 coined truncated unity parquet solver (TUPS). TUPS allows us to reach lower temperatures and finer momenta grids than were possible before either in  fRG or  parquet approaches. We are able to discretize the transfer bosonic momentum ${\vec q}$  with up to $76\times 76$ lattice momenta in the first Brillouin Zone (1.BZ), while at the same time keeping the full frequency dependence of the two-particle vertices.

We  apply TUPS to the two-dimensional Hubbard model and present a
comprehensive analysis of the convergence in the number of form factors for a full-fledged calculation including frequency dependence and self-energy in the diagrammatics. We show that with TUPS it is possible to reach low enough temperatures to reproduce the pseudogap phenomenon as well as a non-violation of the Mermin-Wagner theorem.

Interestingly, the temperature where the pseudogap occurs is sensitive to the number of form factors used, whereas the AFM susceptibility seems to be more sensitive to the ${\vec q}$-resolution.
Nevertheless, we show that in the weak coupling regime, the qualitative description of the 2D Hubbard model with the use of only one form factor is possible and even quantitative differences are small down to low temperatures. In the intermediate to strong coupling regime the quantitative differences between results obtained with only one form factor and with more are stronger.
These differences become even larger, when the parquet approximation is replaced by the D$\Gamma$A method.

 The paper is organized as follows:
In Sec.~\ref{sec:tu_parquet} we begin by reviewing the parquet method in order to make this work self-contained.
We go on to present the TU parquet scheme as derived in Ref.~\onlinecite{Eckhardt18} in our notation. We then also derive a channel unbiased TU Schwinger-Dyson equation.
In Sec.~\ref{sec:numerical_implementation} we discuss our numerical implementation of  TUPS. 
We also describe the memory costs and the computational complexity of the approach.
In Sec.~\ref{sec:qualitative} we present the magnetic susceptibilities and self-energies calculated with TUPS.
In Sec.~\ref{sec:quantitative} we discuss the convergence of calculations with a limited number of basis functions toward the full solution in momentum space. Section \ref{Sec:conclusion} summarizes our results.
In Appendix~\ref{app:su2} we show how to simplify the parquet equation (PE) in the case of SU(2) symmetry.
Afterwards, we derive further optimizations for the PE in Appendix~\ref{app:PE_opt}.
In Appendix~\ref{app:tu_sde} we show the effect a na\"ive implementation of the Schwinger-Dyson equation has on susceptibilities and the pseudogap. Further, in Appendix~\ref{App:D}, we show how the basis truncation influences the crossing symmetry.
Finally, in Appendix~\ref{app:pa_results}, we show additional results omitted in the main text.

\section{\label{sec:tu_parquet}Truncated Unity parquet equations}
\subsection{\label{sec:parquet_recap}Recapitulation of the parquet method}
We start by briefly recapitulating the traditional parquet method as introduced by DeDominicis and Martin\cite{parquetII}  in order to make this work self-contained.
For this we consider a generic, fermionic action relevant for describing condensed matter systems
\begin{equation}
\begin{aligned}
&S[\bar{c}, c] = 
- \frac{1}{2} \int dx dx' \, \bar{c}(x) G_0^{-1}(x, x') c(x') \\
&+ \frac{1}{4} \int dx \cdots dx''' \bar{c}(x) \bar{c}(x') U(x, x', x'', x''')  c(x'') c(x''').
\end{aligned}
\label{eq:fermionic_action}
\end{equation}
Here $\bar{c}$ ($c$) denote creators (annihilators) in the Grassmann algebra and $x$, $x'$, etc.~suitable quantum numbers. 
With the Hubbard model in mind we will use as quantum numbers
\begin{equation}
x = (\vec k, \nu, \sigma)
\label{eq:quantum_numbers}
\end{equation}
where $\vec k$ is a lattice momentum from the first Brillouin zone (1.BZ), $\nu$ an imaginary Matsubara frequency and $\sigma$ the spin.
One can in principle include more quantum numbers as for instance a band index.
With this the partition function $Z$ of the system can be calculated using the fermionic path integral 
\begin{equation}
Z = {\rm const} \times \int D\bar{c} Dc \, e^{-S[\bar{c} , c]}.
\label{eq:partition_function}
\end{equation}
Due to their physical significance, for example for life times and  linear response functions\cite{Kubo57}, we aim at calculating both, the full 1-particle irreducible vertex $\Sigma$ and the  2-particle reducible vertex  $F$ with amputated legs.
$F$ is formally given by a fourth functional derivative of the effective action, which can in turn be obtained as the Legendre-transform of the logarithm of the partition function.\cite{Metzner12, Vasiliev98}
It is possible to show, that diagrams contributing to $F$ can be uniquely classified according to the parquet equation (PE)\cite{parquetII,Bickers04,Rohringer18}
\begin{equation}
F = \Lambda + \sum_{r \in \{\text{ph}, \ov{\text{ph}}, \text{pp}\}} \Phi_{r}.
\label{eq:parquet_no_args}
\end{equation}
where $\Lambda$ denotes the fully 2-particle irreducible vertex and $\Phi_r$ the 2-particle reducible vertices in the particle-hole channel (ph), the transverse particle-hole channel ($\ov{\text{ph}}$) and the particle-particle channel (pp) respectively \cite{Rohringer18,Husemann09,Bickers04}.
$\Lambda$ is considered an input to the parquet scheme.
We further define the diagrams that are 2-particle irreducible in a certain channel as
\begin{equation}
\Gamma_r = F - \Phi_r.
\label{eq:def_gamma_r}
\end{equation}
The equations in the case of a SU(2) symmetric action with explicit arguments are given in Appendix \ref{sec:su2}.

The different 2-particle reducible vertices, on the other hand, can be calculated by means of Bethe-Salpeter equations (BSE)~\cite{Bickers04,Rohringer18}.
For this we first introduce the parametrization of generic 2-particle vertex quantities ($V$) in terms of fermionic arguments as
\begin{equation}
V_{\sigma_1 \sigma_2 \sigma_3 \sigma_4}^{(\vec k_1, \nu_1), (\vec k_2, \nu_2), (\vec k_3, \nu_3), (\vec k_4, \nu_4)} = 
\begin{tikzpicture}[scale = 0.7, every node/.style={scale=0.7}, baseline={([yshift=-.5ex]current bounding box.center)}]
\begin{scope}[very thick,decoration={
  markings,
  mark=at position 0.5 with {\arrow{stealth}}}
  ] 
    \draw [thick] (0.0, 0.0) rectangle (1.0, 1.0);
    \draw node at (0.5, 0.5) {$V$};
    \draw [thick, postaction={decorate}] (-0.5, 1.5) -- (0.0, 1.0);
    \draw node [above] at (-0.5, 1.5) {$(\vec k_1, \nu_1, \sigma_1)$};
    \draw [thick, postaction={decorate}] (1.0, 1.0) -- (1.5, 1.5);
    \draw node [above] at (1.5, 1.5) {$(\vec k_3, \nu_3, \sigma_3)$};
    \draw [thick, postaction={decorate}] (0.0, 0.0) -- (-0.5, -0.5);
    \draw node [below] at (-0.5, -0.5) {$(\vec k_4, \nu_4, \sigma_4)$};
    \draw [thick, postaction={decorate}] (1.5, -0.5) -- (1.0, 0.0);
    \draw node [below] at (1.5, -0.5) {$(\vec k_2, \nu_2, \sigma_2)$};
\end{scope}
\end{tikzpicture}
\label{eq:vertex_fermionic_notation}
\end{equation}
Note that in this equation all legs are amputated.
For time and spatial (within the lattice) translationally invariant systems one can parametrize vertex quantities in terms of three independent lattice momenta and Matsubara frequencies because of energy and momentum conservation. 
Choosing two fermionic arguments $k{:=}(\vec k, \nu)$, $k'{:=}(\vec k', \nu')$ and one bosonic argument $q_r{:=}(\vec q_r, \omega_r)$ in a four-vector notation, one can write the BSE in a compact matrix notation,  introducing vertices as matrices in their fermionic arguments.
Omitting the spin index these read
\begin{equation}
\mat{\Phi}_r^{q_r} = \mat{\Gamma}_r^{q_r} \mat{\chi}_{0,r}^{q_r} \mat{F}^{q_r}
\label{eq:BSE_matrix}
\end{equation}
where  $\mat{\chi}_{0,r}^{q_r}$ denotes the bubble contribution  in terms of the interacting Green's function $G^{k}$ for the respective channel $r$:
\begin{equation}
\begin{aligned}
\mat{\chi}_{0,\text{ph}/\ov{\text{ph}}}^{q_r} &= \left( G^{k} G^{k' + q_r} \delta_{k, k'}%
\right)_{\substack{\vec k \in \text{1.BZ}; n_{\nu} \in \mathbb{Z}\\ \vec k' \in \text{1.BZ}; n_{\nu'} \in \mathbb{Z}}}\\
\mat{\chi}_{0,\text{pp}}^{q_r} &= \left( G^{k} G^{q_r - k'} \delta_{k, k'}%
\right)_{\substack{\vec k \in \text{1.BZ}; n_{\nu} \in \mathbb{Z}\\ \vec k' \in \text{1.BZ}; n_{\nu'} \in \mathbb{Z}}}.
\end{aligned}
\label{eq:def_chi_matrix}
\end{equation}
That is, the $\chi$'s are diagonal matrices in the combined index of lattice momentum and frequency.
In order for these equations to hold one needs to chose $(\vec q_r, \omega_r)$ to be the respective transfer/total momentum/frequency in the different channels: $k_4 - k_1$ in case of ph, $k_3 - k_1$ in case of $\ov{\text{ph}}$ and $k_1 + k_2$ in case of pp.
This channel specific parametrization of vertex quantities will be called 'channel native' parametrization  in the following, and the corresponding bosonic momentum/frequency 'channel native' momentum/frequency.

Another quantity we are interested in is the one-particle irreducible single-particle vertex aka fermionic self-energy $\Sigma$.
Having the full $F$ available, it can be calculated by means of the Schwinger-Dyson equation (SDE)
\begin{equation}
\begin{aligned}
\Sigma^{k}_{\sigma}& = \frac{1}{ \beta N} \sum_{k',\sigma_1} G^{k'}_{\sigma_1} U^{k', k, q=0}_{\sigma, \sigma_1, \sigma, \sigma_1}\\
+ &\frac{1}{2 \beta^{2} N^2} \sum_{k', q} \sum_{\sigma_1, \sigma_2, \sigma_3} 
 F^{k, k', q}_{\sigma, \sigma_1, \sigma_2, \sigma_3} G^{k' + q}_{\sigma_1} G^{k}_{\sigma_2} G^{k + q}_{\sigma_3} U^{k', k, q}_{\sigma_2, \sigma_3, \sigma, \sigma_1}.
\end{aligned}
\label{eq:SDE}
\end{equation}
The above equations (\ref{eq:parquet_no_args}), (\ref{eq:def_gamma_r}), (\ref{eq:BSE_matrix}) and (\ref{eq:SDE}), also often referred together as the PEs, are exact and allow for calculating $F$, $\Phi_r$ , $\Gamma_r$, and  $\Sigma$, if the fully irreducible vertex $\Lambda$ is given. 
But their mutual dependence is complex.
The calculation of $F$ through the PE Eq.~\eqref{eq:parquet_no_args} requires knowledge of the different $\Phi$ and $\Lambda$, the calculation of $\Phi_r$ through the BSE~\eqref{eq:BSE_matrix} requires the knowledge of the Green's function or equivalently the self-energy $\Sigma$ and the calculation of $\Sigma$ with the SDE~\eqref{eq:SDE} directly requires an expression for $F$.
Thus one way to find a solution to the above set of equations is to iterate them from a known starting point to a fix-point which then represents the sought after solution (for an alternative approach to find a fixed point via differential equations see Ref.~\onlinecite{Kugler18}).

\subsubsection{\label{sec:lambda}Parquet approximation and  D$\Gamma$A}
As an exact expression for $\Lambda$ is not known for any but the most trivial models, one further needs to  come up with approximations.
One possibility is to calculate $\Lambda$ in lowest order perturbation theory which amounts to setting $\Lambda = U$, i.e.,  the bare interaction from Eq.~\eqref{eq:fermionic_action} which amounts to the so-called parquet approximation (PA). 
Another possibility is to include all local diagrams in $\Lambda$ by solving an auxiliary Anderson impurity model. For the local Green's function e.g. a converged DMFT solution is taken.
 This is the D$\Gamma$A approach\cite{Toschi2007,Katanin2009,Rohringer18} which besides the local correlations of DMFT also includes non-local correlations, e.g.~antiferromagentic spin fluctuations, pseudogaps\cite{Katanin2009,Schaefer2015-3}, (quantum) critical behavior\cite{Rohringer2011,Schaefer2016,DelRe2018,Schaefer2019} and superconductivity\cite{Kitatani2019,Kauch2019}. 
In a more elaborate calculation, one can also require that the local part of the D$\Gamma$A Green's function coincides with the local Green's function of the auxiliary impurity model. This amounts to an additional update of the impurity problem. This step has not been performed in this work.~\footnote{Preliminary studies for the 2D Hubbard model for the parameters here considered show that this update does not influence the results notably.}  
A closely related approach is the dual fermion approach\cite{Rubtsov2008} which takes the full vertex $F$ from an impurity model as a starting point and connects these building blocks by non-local Green's function contributions, either by ladder diagrams\cite{Hafermann2009} or the PEs\cite{Astleithner2019}.

\subsection{\label{sec:intro_TU}Introduction of the Truncated Unity scheme}
In this Section, we introduce the truncated unity (TU) method for the parquet equations.
We here give the derivations for a generic action relevant for many problems related to condensed matter physics. Out notation and basis is different from  Ref.~\onlinecite{Eckhardt18}, where the  transformed BSE and PE but not the transformed SDE have already been introduced.

Let us consider the 2-particle reducible vertices $\Phi_{r}$ parametrized in terms of two fermionic arguments ($k$ and $k'$) and their respective `channel native' transfer/total momentum/frequency $q_r$, as introduced above.
In what follows we will omit spin arguments and sums since all derivations are completely independent of spin. In the presence of SU(2) symmetry also often even/odd or density/magnetic spin combinations are employed \cite{RMPVertex,Rohringer2018}, and in Appendix \ref{App:A} we briefly discuss how to extend the equations if these  spin combinations are employed.

The TU approximation is based on the idea that the 2-particle reducible vertices $\Phi_r$ are more local in their two fermionic arguments in real space than in their respective bosonic argument\cite{Husemann09,Wang2012a}.
This has for example previously been seen to hold for generic situations by systematic explorations of the basis length for the 2D Hubbard model in fRG\cite{Lichtenstein17} and has also  been observed when solving the PEs~\cite{Kauch2019,Pudleiner2019a}. The huge difference is also most obvious when approaching a phase transition, as here the susceptibility  evolves towards a $\delta$-peak in the  respective bosonic $\vec q$ momentum.

Conversely, this means that the dependence on the fermionic lattice momenta $\vec k$, $\vec k'$ is weak  compared to the dependence on $\vec q_r$.
We, therefore, try to exploit this in order to achieve a dimensional reduction of the vertex quantities by introducing a new basis
\begin{equation}
f:\mathbb{N}\times \text{1.BZ} \rightarrow \mathbb{C}; \hspace{3mm} (\ell, \vec k) \rightarrow f^{\ell, \vec k}
\label{eq:def_basis_functions}
\end{equation}
in which we represent the vertex functions.
We call this new basis the form-factor basis (in accordance with the term used frequently in fRG literature \cite{Lichtenstein17,Tagliavini19}).
Here $\vec k$ denotes a lattice momentum in the first Brillouin Zone (1.BZ) and $\ell$ labels functions in the new basis.
We ask for the following three conditions of the basis functions:
(i) they should be complete and orthogonal in the sense that
\begin{equation}
\begin{aligned}
&\sum_{\ell} f^{\ell, \vec k} \left( f^{\ell, \vec k'} \right)^{*} = N \delta_{\vec k, \vec k'}\\
\frac{1}{N} & \sum_{\vec k} f^{\ell, \vec k} \left( f^{\ell', \vec k} \right)^{*} = \delta_{\ell, \ell'}.
\end{aligned}
\label{eq:orthonormality_completeness}
\end{equation}
(ii) it should be a real space basis such that each basis function is uniquely assignable to a distance in the real lattice,
(iii) one may further ask for a specific transformation behavior or symmetries of the functions.
Conditions (i) and (ii) are readily fulfilled by using the simple real-space basis of Ref.~\onlinecite{Eckhardt18}. 

Here,  in order to also fulfill (iii), we  take suitable linear combinations of complex exponentials as explained in Ref.~\onlinecite{Platt13} with the 2D square and the hexagonal lattice as specific examples.
For the example of the square lattice, defining functions  that transform according to the irreducible representations of the point group and choosing prefactors such that the functions are real, one  gets as the first few basis functions:
\begin{equation}
\begin{aligned}
f^{1, \vec k} &= 1\\
f^{2, \vec k} &= \frac{1}{2} (\cos(k_x) + \cos(k_y))\\
f^{3, \vec k} &= \frac{1}{2} (\cos(k_x) - \cos(k_y))\\
f^{4, \vec k} &= \sin(k_x)\\
f^{5, \vec k} &= \sin(k_y).
\end{aligned}
\label{eq:examples_basis}
\end{equation}
With this basis at hand we define the 2-particle reducible vertices in the form-factor basis as 
\begin{equation}
\begin{aligned}
&\tilde{\Phi}_r^{(\ell_1, \nu_1), (\ell_2, \nu_2), (\vec q_r, \omega_r)} := \\
& \sum_{\vec k_1, \vec k_2 \in 1. \text{BZ}} \hspace{-2mm} f^{\ell_1, \vec k_1} 
\Phi_r^{(\vec k_1, \nu_1), (\vec k_2, \nu_2), (\vec q_r, \omega_r)} (f^{\ell_2, \vec k_2})^{*}.
\end{aligned}
\label{eq:def_trafo_vertex}
\end{equation}
As stated before, the TU-approximation relies on the vertices $\Phi_r$ to decay quickly in their fermionic real-space arguments.
Expressed more formally one could write, omitting frequencies
\begin{equation}
\left|\tilde{\Phi}_r^{\ell_1, \ell_2, q_r} \right|_{|\ell_1 > \ell_{\text{max}} \lor \ell_2 > \ell_{\text{max}}} \ll %
\left|\tilde{\Phi}_r^{\ell_1, \ell_2, q_r} \right|_{|\ell_1 \leq \ell_{\text{max}} \land \ell_2 \leq \ell_{\text{max}}}
\label{eq:expansion_parameter}
\end{equation}
for some value $\ell_{\text{max}}$ corresponding to a certain distance in the real lattice $d_{\ell_{\text{max}}}$.
We aim at using Eq.~\eqref{eq:expansion_parameter} to justify saving vertices only for a reduced set of basis functions using the fact that vertices for $\ell > \ell_{\text{max}}$ are small compared to vertices with $\ell \leq \ell_{\text{max}}$ in order to adjust and simplify calculations. This will lead to considerably reduced memory and computing time requirements.
In the following we show how this can be done in the BSE, the PE and the SDE.
%
%
\subsection{Bethe-Salpeter equation}
Consider the BSE for channel $r$ in a matrix notation as given in Eq.~\eqref{eq:BSE_matrix}.
We introduce the transformation matrix
\begin{equation}
\mat{U} = \left( U^{(\ell, \nu)(\vec k, \nu')} \right)_{\substack{\ell \in \mathbb{N}; {n_{\nu}} \in \mathbb{Z}\\ \vec k \in \text{1.BZ}; n_{\nu'} \in \mathbb{Z}}} = \frac{1}{\sqrt{N}} f^{\ell, \vec k} \delta_{\nu, \nu'}
\label{eq:def_U_matrix}
\end{equation}
which besides the $f^{\ell, \vec k}$ now also includes the Matsubara frequencies $\nu$ and $\nu'$.   From Eq.~\eqref{eq:orthonormality_completeness} and the $\delta$-function in Matsubara frequencies we directly get
\begin{equation}
\mat{U} \, \mat{U}^{\dagger} = \mat{U}^{\dagger} \mat{U} = I \; ,
\label{eq:unitarity}
\end{equation}
i.e., the matrix $\mat{U}$ is unitary.
With this we can write, denoting with a tilde the corresponding quantities in the new, form-factor basis:
\begin{equation}
\begin{aligned}
\tilde{\mat{\Phi}}_r^{q_r} = &\mat{U} \, \mat{\Gamma}_r^{q_r} \mat{\chi}_{0,r}^{q_r} \mat{F}^{q_r} \mat{U}^{\dagger}\\
=&\mat{U} \, \mat{\Gamma}_r^{q_r} \mat{U}^{\dagger} \mat{U} \, \mat{\chi}_{0,r}^{q_r} \mat{U}^{\dagger} \mat{U} \, \mat{F}^{q_r} \mat{U}^{\dagger}\\
=& \tilde{\Gamma}_{r}^{q_r} \tilde{\chi}_{0,r}^{q_r} \tilde{F}_{r}^{q_r} \; .
\end{aligned} 
\label{eq:transformed_BSE}
\end{equation}
Using the full form-factor basis this is simply the basis transform of a chain of matrix products. \red{In the truncated basis the matrix $\mat{U}$ does not have an inverse and Eq.~\eqref{eq:unitarity} does not hold. In the following we will derive the expressions for all vertex quantities in the full form-factor basis, introducing the basis truncation later in the already derived expressions.
}
 
From Eq.~\eqref{eq:parquet_no_args} we see that in order to express all occurring vertex quantities in terms of $\Lambda$ and 2-particle reducible vertices we need the quantities
\begin{equation}
\begin{aligned}
\tilde{\mat{\Lambda}}^{q_r} &:= \mat{U} \, \mat{\Lambda}^{q_r} \, \mat{U}^{\dagger}\\
\tilde{\mat{\Phi}}_r^{q_r} &:= \mat{U} \, \mat{\Phi}_r^{q_r} \, \mat{U}^{\dagger}\\
\tilde{\mat{\Phi}}_h^{q_r} &:= \mat{U} \, \mat{\Phi}_h^{q_r} \, \mat{U}^{\dagger}; \hspace{2mm} h \neq r.
\end{aligned}
\label{eq:quantities for BSE}
\end{equation}
The last equation differs from the previous in that the vertices in channel $h$ are here parametrized in the channel native parametrization of channel $r$.
For given $\Lambda$, $\tilde{\mat{\Lambda}}$ is trivially known.
$\tilde{\mat{\Phi}}_r$ is the same as on the LHS of Eq.~\eqref{eq:transformed_BSE} and will thus not be further expressed by means of other equations.
It, therefore, remains to be shown how to obtain the quantity $\tilde{\mat{\Phi}}_h$, $h \neq$ r which we will do by means of a  transformation of the PEs.
\subsection{\label{sec:derivation_tu_parquet}Parquet equation}
Let us consider the PEs as given in the spin-diagonal notation in Eq.~\eqref{eq:parquet_SU2}. Considering the frequency and momentum arguments in the round brackets, there are only four distinct  frequency and momentum combinations  in Eq.~\eqref{eq:parquet_SU2}.
We will discuss the transformation of the first term only - the others then follow in exactly the same way.
An analogous derivation in a fRG context can be found in Ref.~\onlinecite{Tagliavini19}.

To be specific we consider the first nontrivial contribution to the irreducible vertex $\Gamma_d$ in Eq.~\eqref{eq:parquet_SU2}. It is that of the $\overline{\text{ph}}$-channel to the ph-channel (here and in the following we denote this contribution with an arrow: \mbox{ph $\leftarrow \overline{\text{ph}}$}). This contribution reads explicitly
\begin{widetext}
\begin{equation}
\begin{aligned}
\Gamma_{d}&^{(\vec k_1, \nu_1)(\vec k_2, \nu_2)(\vec q_1, \omega_1)} \leftarrow 
\left[ - \frac{1}{2} \Phi_d - \frac{3}{2} \Phi_m \right]^{(\vec k_1, \nu_1)(\vec k_1 + \vec q_1, \nu_1 + \omega_1)(\vec k_2 - \vec k_1, \nu_2 - \nu_1)}.
\end{aligned}
\label{eq:first_contrib_parquet}
\end{equation}
We start by relabeling the arguments ($q_2:=k_2-k_1$) in a convenient way, expressing the last argument on the right hand side also explicitly as a bosonic argument: 
\begin{equation}
\begin{aligned}
\Gamma_{d}&^{(\vec k_1, \nu_1)(\vec k_1 + \vec q_2, \omega_2 + \nu_1)(\vec q_1, \omega_1)} \leftarrow 
\left[ - \frac{1}{2} \Phi_d - \frac{3}{2} \Phi_m \right]^{(\vec k_1, \nu_1)(\vec k_1 + \vec q_1, \nu_1 + \omega_1)(\vec q_2, \omega_2)}.
\end{aligned}
\label{eq:first_contrib_parquet_new_arguments}
\end{equation}
We can now use the basis transformation in Eq.~\eqref{eq:def_trafo_vertex} to compute the quantities needed in the BSE~\eqref{eq:transformed_BSE}
\begin{equation}
\begin{aligned}
\tilde{\Gamma}_{d}&^{(\ell_1, \nu_1), (\ell_2, \omega_2 + \nu_1), (\vec q_1, \omega_1) }
=
\sum_{\vec k_1 \vec q_2} f^{\ell_1, \vec k_1} 
\Gamma_{d}^{(\vec k_1, \nu_1), (\vec q_2 + \vec k_1, \omega_2 + \nu_1), (\vec q_1, \omega_1)}
f^{\ell_2, \vec k_1 + \vec q_2} \\
&\leftarrow
\sum_{\vec k_1 \vec q_2} f^{\ell_1, \vec k_1}
\left[ -\frac{1}{2} \Phi_{d} - \frac{3}{2} \Phi_{m} \right]^{(\vec k_1, \nu_1), (\vec q_1 + \vec k_1, \omega_1 + \nu_1), (\vec q_2, \omega_2)}
f^{\ell_2, \vec k_1 + \vec q_2} \\
&=
\sum_{\vec k_1 \vec q_2} f^{\ell_1, \vec k_1} f^{\ell_2, \vec k_1 + \vec q_2}  
\sum_{\ell_3 \ell_4} f^{\ell_3, \vec k_1} f^{\ell_4, \vec k_1 + \vec q_1}
\left[ -\frac{1}{2} {\tilde{\Phi}}_{d} - \frac{3}{2} {\tilde{\Phi}}_{m} \right]^{(\ell_3, \nu_1), (\ell_4, \omega_1 + \nu_1), (\vec q_2, \omega_2)}\\
&=
\sum_{\ell_3 \ell_4} \sum_{\vec q_2} 
\left[ -\frac{1}{2} {\tilde{\Phi}}_{d} - \frac{3}{2} {\tilde{\Phi}}_{m} \right]^{(\ell_3, \nu_1), (\ell_4, \omega_1 + \nu_1), (\vec q_2, \omega_2)}
\sum_{\vec k_1} f^{\ell_1, \vec k_1} f^{\ell_2, \vec k_1 + \vec q_2}
f^{\ell_3, \vec k_1} f^{\ell_4, \vec k_1 + \vec q_1}.\\
\end{aligned}
\label{eq:trafo_PE}
\end{equation}
\end{widetext}
The turning around of arguments to insert a vertex from channel $h$ into channel $r$ is achieved here by transforming $\tilde{\Phi}_r$ back to $\vec k$-space and then expanding it via the form-factor expansion in channel $r$.
This operation, however, poses a problem when performing calculations in a reduced basis set.\cite{Eckhardt18}
The reason for this is that the transformation back to $\vec k$-space is then only  possible in an approximate way.
Still performing the above steps also in a reduced basis introduces the extra approximation that the 2-particle reducible vertices couple into the other channels with a bosonic argument that is approximated in a similar way as the respective fermionic arguments.
This is not compatible with our initial intent to approximate only the fermionic arguments keeping the full bosonic momentum dependence.
However, the fact that the bosonic argument is inevitably approximated in the channel cross-insertions also gives room for a further optimization of the equations leading to a PE scheme  that scales linearly in the number of discrete lattice momenta in the 1.BZ.
For an explicit calculation showing this see Appendix~\ref{app:PE_opt}.
 
The fact that the cross-inserted vertices with a strong dependence on the transfer momentum are harder to capture with a few basis functions indicates that cross-insertions of  long-range interactions, either due to unscreened Coulomb interactions or generated by soft modes, might get truncated when they feed back into other channels. For instance, theoretically, a critical magnetic mode could trigger very long-range Cooper pairs. These would not be seen by this truncated basis approach.

\subsection{\label{sec:derivation_tu_sde}Schwinger-Dyson equation}
We start by rewriting the SDE~\eqref{eq:SDE} such that each component of $F$ ($\Lambda$, $\Phi_r$) appears in its channel native parametrization. {To simplify notation we restrict the bare interaction vertex $U^{k,k',q}_{\sigma_1, \sigma_2, \sigma_3, \sigma_4}$ to the bare Coulomb interaction of the Hubbard model (see Sec.~\ref{sec:HM}).} 
The SDE then reads
\begin{equation}
\begin{aligned}
&\underline \Sigma = { \frac{Un}{2}}\\
& - {U}\sum_{\vec q_1, \omega_1} \left[ \mat{\Lambda}^{(\vec q_1, \omega_1)} \otimes \underline \chi_{0, \text{ph}}^{(\vec q_1, \omega_1)} \right] \cdot \underline G_{\text{ph}}^{(\vec q_1, \omega_1)}\\
& - \frac{{U}}{2} \sum_{\vec q_1, \omega_1} \left[ \mat{\left[ \Phi_d - \Phi_m \right]}^{(\vec q_1, \omega_1)} \otimes \underline \chi_{0, \text{ph}}^{(\vec q_1, \omega_1)} \right] \cdot\underline G_{\text{ph}}^{(\vec q_1, \omega_1)}\\
& + {U}\sum_{\vec q_1, \omega_1} \left[ \mat{\Phi}_m^{(\vec q_1, \omega_1)} \otimes \underline \chi_{0, \text{ph}}^{(\vec q_1, \omega_1)} \right]\cdot \underline G_{\text{ph}}^{(\vec q_1, \omega_1)}\\
& - \frac{{U}}{2} \sum_{\vec q_1, \omega_1} \left[ \mat{\left[ \Phi_s - \Phi_t \right]}^{(\vec q_1, \omega_1)} \otimes \underline \chi_{0, \text{pp}}^{(\vec q_1, \omega_1)} \right] \cdot\underline  G_{\text{pp}}^{(\vec q_1, \omega_1)}.
\end{aligned}
\label{eq:SDE_split_up}
\end{equation}
Here we have defined
\begin{equation}
\begin{aligned}
\underline \Sigma = \left( \Sigma^{\vec k, \nu} \right)_{\vec k \in \text{1.BZ};n_{\nu}\in \mathbb{Z}}&\\
\underline \chi_{0, \text{ph}}^{(\vec q, \omega)} = \left( \chi_{0, \text{ph}}^{(\vec k, \nu)(\vec q, \omega)} \right)_{\vec k \in \text{1.BZ}; n_{\nu} \in \mathbb{Z}} &:=%
G(\vec k {+} \vec q, \nu {+} \omega) G(\vec k, \vec \nu)\\
\underline \chi_{0, \text{pp}}^{(\vec q, \omega)} = \left( \chi_{0, \text{pp}}^{(\vec k, \nu)(\vec q, \omega)} \right)_{\vec k \in \text{1.BZ}; n_{\nu} \in \mathbb{Z}} &:=%
G(\vec q {-} \vec k, \omega {-} \nu) G(\vec k, \vec \nu)\\
\underline G_{\text{ph}}^{(\vec q, \omega)} = \left( G_{\text{ph}}^{(\vec k, \nu)(\vec q, \omega)} \right)_{\vec k \in \text{1.BZ}; n_{\nu} \in \mathbb{Z}} &:=%
G(\vec k {+} \vec q, \nu {+} \omega)\\
\underline G_{\text{pp}}^{(\vec q, \omega)} = \left( G_{\text{pp}}^{(\vec k, \nu)(\vec q, \omega)} \right)_{\vec k \in \text{1.BZ}; n_{\nu} \in \mathbb{Z}} &:=%
G(\vec q {-} \vec k, \omega {-} \nu) \; ;
\end{aligned}
\label{eq:def_greens_quantities}
\end{equation}
$\otimes$ denotes the ordinary matrix vector product and $\cdot$ the component-wise multiplication of the vectors.
Since it is our goal to keep the self-energy in $\vec k$-space all that remains to be done is to express the occurring vertices in terms of their form-factor expansions.
Since we assume $\Lambda$ to be given, it can easily be expressed in the form-factor basis.
The other contributions are already conveniently written in their respective matrix form making the transformations easy.
We can thus directly write down the transformed SDE as:\\
\begin{equation}
\begin{aligned}
&\underline \Sigma = { \frac{Un}{2}}\\
& - {U}\sum_{\vec q_1, \omega_1} \left[ \mat{U}^{\dagger} \tilde{\mat{\Lambda}}^{(\vec q_1, \omega_1)} \mat{U} \otimes \underline \chi_{0, \text{ph}}^{(\vec q_1, \omega_1)} \right] \cdot \underline G_{\text{ph}}^{(\vec q_1, \omega_1)}\\
& - \frac{{U}}{2} \sum_{\vec q_1, \omega_1} \left[ \mat{U}^{\dagger} \mat{\left[ \tilde{\Phi}_d - \tilde{\Phi}_m \right]}^{(\vec q_1, \omega_1)} \mat{U} \otimes \underline \chi_{0, \text{ph}}^{(\vec q_1, \omega_1)} \right]  \cdot \underline G_{\text{ph}}^{(\vec q_1, \omega_1)}\\
& + {U}\sum_{\vec q_1, \omega_1} \left[ \mat{U}^{\dagger} \tilde{\mat{\Phi}}_m^{(\vec q_1, \omega_1)} \mat{U} \otimes \underline \chi_{0, \text{ph}}^{(\vec q_1, \omega_1)} \right] \cdot  \underline G_{\text{ph}}^{(\vec q_1, \omega_1)}\\
& - \frac{{U}}{2} \sum_{\vec q_1, \omega_1} \left[ \mat{U}^{\dagger} \mat{\left[ \tilde{\Phi}_s - \tilde{\Phi}_t \right]}^{(\vec q_1, \omega_1)} \mat{U} \otimes \underline \chi_{0, \text{pp}}^{(\vec q_1, \omega_1)} \right] \cdot  \underline G_{\text{pp}}^{(\vec q_1, \omega_1)}.
\end{aligned}
\label{eq:SDE_transformed}
\end{equation}

\section{\label{sec:numerical_implementation}Numerical implementation for the 2D Hubbard model}

\subsection{\label{sec:HM}Model and parameters used}
As a test ground for our method we use the paradigmatic 2D Hubbard model with nearest-neighbour hopping $t$, local interaction $U$ and a single band on a square lattice.
For this model the Hamiltonian reads
\begin{equation}
H = -t \sum_{\langle i, j \rangle, \sigma}   c^{\dagger}_{i, \sigma} c_{j, \sigma} + \mu \sum_{i, \sigma}  n_{i, \sigma}%
+ U \sum_{i} n_{i, \uparrow} n_{j, \downarrow}.
\label{eq:Hubbard_Hamiltonian}
\end{equation}
Here,  $c^{\dagger}_{i, \sigma}$ and  $c_{j, \sigma}$ are the creation and annihilation operator, respectively,  for a particle with spin $\sigma$ on site $i$.
In the following sections we restrict ourselves to half-filling, $n=1$, and two values of the local Coulomb interaction: $U=2t$ and $U=4t$. In the following, the hopping $t\equiv 1$, $\hbar\equiv 1$, $k_B\equiv 1$, and lattice constant $a\equiv 1$ set the units of energy, temperature, and distance, respectively.

\subsection{Approximations for the fully irreducible vertex}

The parquet equations are solved in the TUPS implementation, which is made available at \onlinecite{TUPS}, using two different approximations for the fully irreducible vertex $\Lambda$ (see also Sec.~\ref{sec:lambda}):
\begin{itemize}
\item[(i)] the parquet approximation (PA)~\cite{Bickers04} where the fully irreducible vertex is taken in lowest order: $\Lambda=U$; 
\item[(ii)] the parquet D$\Gamma$A~\cite{Valli2015,Li2016,Kauch2019}, where $\Lambda$ is taken from a converged DMFT calculation, i.e. $\Lambda$ is local and approximated by a fully irreducible vertex of an auxiliary impurity problem. The impurity problem was solved with the continuous time quantum Monte-Carlo method using the {\it w2dynamics}~\cite{w2dynamics} package. This way  $\Lambda$ consists of all local diagrams that are built from the local DMFT Green's function lines and the local interaction $U$. 
\end{itemize}

\subsection{Memory and computational costs} 

Due to the TU approximation memory costs for calculating vertex functions now scale like $\mathcal{O}(N_{\text{FF}}^{2} N_{q} N_{\omega}^{3})$ compared to $\mathcal{O}(N_{q}^{3} N_{\omega}^{3})$ for the plain-vanilla parquet implementation.%
\footnote{It is valid to consider the scaling with $N_{\text{FF}}$ and $N_{q}$ separately since we have checked that the convergence in the number of basis functions $N_{\text{FF}}$ does not strongly depend on the number of discrete lattice momenta taken into account $N_{q}$.}
This practically reduces memory costs drastically due to the need to take only few basis functions $N_{\text{FF}}\ll N_{q}$ in the case of the Hubbard model.
For an illustration we compare the memory consumption of a single vertex function in the $\vec k$-space parquet scheme and the TU-scheme in Table \ref{tab:memory_tu} and Table \ref{tab:memory_pa}. For this we assume typical system sizes and a TU calculation using $N_{\text{FF}}=9$ basis functions. 
We will argue below that this is a reasonable amount.
\begin{table}[h]
\centering
\begin{tabular}{ c | c c c c c c} 
 $N_{\omega}$/$N_{k}$ & $4 \times 4$ & $8 \times 8$ & $12 \times 12$ & $16 \times 16$ & $20 \times 20$ & $56 \times 56$ \\
\hline 
 $32$ & $0.68$ & $2.72$ & $6.12$ & $10.87$ & $16.99$ & $133.18$ \\ 
 $48$ & $2.29$ & $9.17$ & $20.64$ & $36.69$ & $57.33$ & $449.47$ \\ 
 $64$ & $5.44$ & $21.74$ & $48.92$ & $86.77$ & $135.90$ & $1065.42$ \\ 
 $80$ & $10.62$ & $42.46$ & $95.55$ & $169.87$ & $265.42$ & $2080.90$ \\ 
 \hline
\end{tabular}
\caption{Memory cost of a single vertex function in GB in the TU-parquet scheme for different system sizes and $9$ basis functions.
The last entry is the largest system size we were able to simulate within reasonable computational cost.
\label{tab:memory_tu}}
\vskip 0.2cm
\centering
\begin{tabular}{ c | c c c c c c} 
 $N_{\omega}$/$N_{k}$ & $4 \times 4$ & $8 \times 8$ & $12 \times 12$ & $16 \times 16$ & $20 \times 20$ & $56 \times 56$ \\
\hline 
 $32$ & $2$ & $137$ & $1566$ & $8797$ & $33554$ & $16169555$ \\ 
 $48$ & $7$ & $464$ & $5284$ & $29687$ & $113246$ & $54572249$ \\ 
 $64$ & $17$ & $1100$ & $12524$ & $70369$ & $268435$ & $129356443$ \\ 
 $80$ & $34$ & $2147$ & $24461$ & $137439$ & $524288$ & $252649304$ \\ 
 \hline
\end{tabular}
\caption{Same as Table \ref{tab:memory_tu} but for the conventional $\vec k$-space parquet scheme.
\label{tab:memory_pa}}
\end{table}%

{In addition to the reduced memory costs} the computational complexity of performing a TUPS calculation is also greatly reduced when compared to the original $\vec k$-space scheme.
The heavy computational task for solving the PEs mainly consists of evaluating BSE, actual PE and SDE.
The evaluation of the BSE now scales like $\mathcal{O}(N_{\text{FF}}^{3}N_{q}N_{\omega}^{4})$ compared to $\mathcal·{O}(N_{q}^{4} N_{\omega}^{4})$ in the original $\vec k$-space scheme.
For analyzing the complexity of the PE we considered the optimized form from  Appendix \ref{app:PE_opt}.
It can be evaluated with $\mathcal{O}(N_{\text{FF}}^{5} N_{q} N_{\omega}^{3} + N_{q}^{2} N_{\text{FF}}^{4})$ computations.
The scaling with $\mathcal{O}(N_{q}^{2})$ stems from the calculation of projection matrices that can be precalculated before the actual iteration and saved without high memory costs. Also the factor of $N_{w}^{3}$ (number of Matsubara frequencies) of the first term in many practical cases more than compensates for the additional factor of $N_{q}$ in the second term.
In the case of many iteration steps the amortized costs for the evaluation of the PE are therefore essentially given by $\mathcal{O}(N_{\text{FF}}^{5} N_{q} N_{\omega}^{3})$.%
With these considerations the scaling of the BSE and the PE are in particular linear in the number of discrete lattice momenta $N_{q}$.

The SDE can be evaluated, saving several intermediate results that cost less memory than the vertex functions, with a computational cost of
$\mathcal{O}(N_{\text{FF}}^{2} N_{q}^{2} N_{\omega}^{2} + N_{\text{FF}}^{2} N_{q} N_{\omega}^{3})$. These are huge savings in floating point operations since $N_{\text{FF}}$ may be kept relatively small\footnote{{A possible computational advantage/disadvantage might also be a faster/slower convergence of the self-consistent parquet equations. However, in our experience the convergence rate  remains very similar when using form factors.}}. 
In the TUPS implementation, thanks to the smaller vertex sizes, the computation time and the on-node memory access are the main constraints whereas the conventional parquet solution was limited by available memory and internode communication due to access to distributed memory.

\subsection{TUPS implementation}
We have realized an implementation of the TU parquet method with the TUPS (\underline{T}runcated \underline{U}nity \underline{P}arquet \underline{S}olver).
This program is parallelized using the combined MPI and OpenMP model.
The MPI parallelization scheme is based  on that of the {\it victory} code~\cite{victory} which was originally proposed in Ref.~\onlinecite{Jarrell13}.
To treat the frequency dependence of the vertices we use an asymptotics scheme similar to that proposed in  Ref.~\onlinecite{Li16}. Also a finer momentum grid was used whenever computationally cheap, i.e., where simple products of Green's functions occurred in the SDE and BSE (for details see Ref.~\onlinecite{Kauch2019}). For this finer momentum grid seven times more $\vec k$-points in each direction have been taken, which was sufficient for convergence.

\section{\label{sec:qualitative}Results for the half-filled Hubbard model}
In this part we demonstrate that with our method we are able to reach low temperatures where the physics qualitatively differs form mean-field behavior.
We will show and explain that it is possible to obtain the same qualitative results already with a single basis function within our form-factor expansion, at least in the parameter regime studied where spin fluctuation physics dominates.

The reason we are able to adequately describe this regime is that our method enables us to discretized the bosonic argument of different vertex quantities with up to several thousand discrete momenta in the 1.BZ.
This is necessary to obtain reliable results due to the onset of longer-ranged AFM correlations at low temperatures, reminiscent of the AFM ground state at half filling for arbitrary small interactions \cite{Schaefer2015-2}.
These correlations show up in the bosonic argument of vertex quantities and not in the fermionic arguments, due to the here chosen parametrization.
Therefore, one can reproduce the non-mean-field behavior already with a single basis function within our form-factor expansion.

\subsection{Magnetic susceptibility\label{sec:inverse_susceptibility}}
\begin{figure}[tb]
\centering
\includegraphics{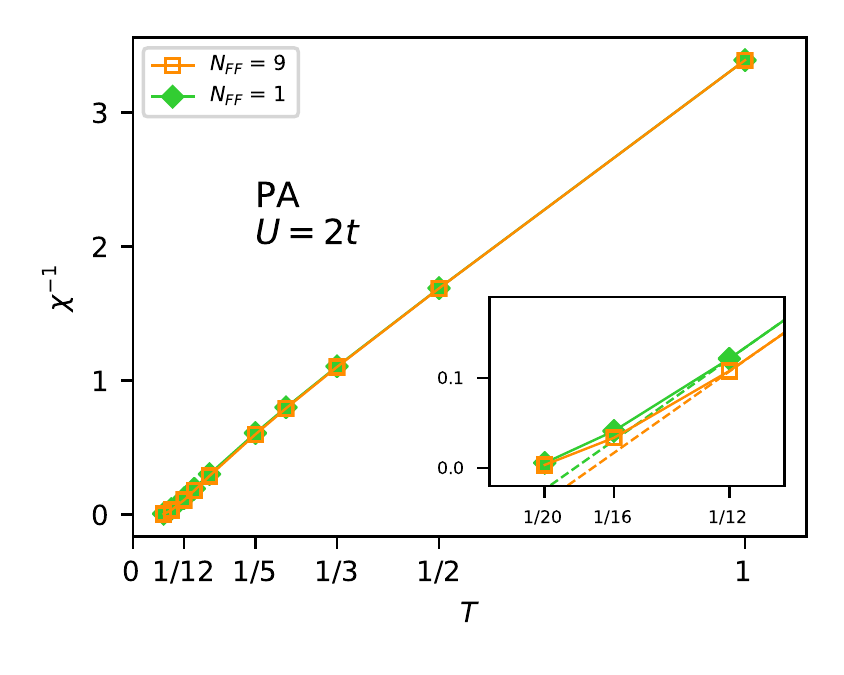}
\vspace{-5mm}
\caption{
Inverse magnetic susceptibility $\chi_m(\vec q = (\pi, \pi), \omega = 0)^{-1}$ of the half-filled Hubbard model as a function of temperature $T$ at $U=2t$ in the PA for $N_{\text{FF}}=1$ and $N_{\text{FF}}=9$.
Inset: zoom into the low temperature region.
Dotted lines in the inset are a linear fit to the respective inverse susceptibilities at $T=1/8$, $1/10$, $1/12$. 
}
\label{fig:chi_inv}
\end{figure}
We begin by discussing the magnetic susceptibility $\chi_m$ at $\vec q = (\pi, \pi)$ and $\omega = 0$, i.e., the AFM susceptibility, the inverse of which we plot for different temperatures in Fig.~\ref{fig:chi_inv}. The calculations have been performed at weak coupling ($U = 2t$) using the PA which is known to yield reliable results in this parameter regime~\cite{Thomas_paper}. The two different curves correspond to different number of form factors used: $N_{\text{FF}}=1$ and $N_{\text{FF}}=9$.  The results are converged in the number of discrete lattice momenta and Matsubara frequencies down to $\beta = 16/t$.
For the point at $\beta = 20/t$ we extrapolated to infinite momentum-grid size.

At high $T$ we obtain a Curie-Weiss like behavior of the inverse susceptibility. When lowering the temperature we begin to see deviations from this for both calculations with a single basis function and with $N_{\text{FF}}=9$ basis functions. In the inset of Fig.~\ref{fig:chi_inv} the linear  (mean-field) fits to the low temperature data ($\beta = 8/t$, $10/t$ and $12/t$) are shown with dotted lines. For $\beta = 16/t$ and $\beta = 20/t$ the inverse susceptibilities lie above the linear  fit. This hints that the PA respects the Mermin-Wagner theorem~\cite{Mermin1966}, which prohibits the breaking of continuous symmetries, i.e. , also AFM order, at finite temperatures in two dimensions. Previous studies within the PA or with the multi-loop fRG (mfRG)~\cite{Tagliavini19,Vilardi19} could not achieve low enough temperatures to see this 'upturn' of inverse susceptibility, either due to insufficient momentum discretization or lack of self-energy feedback. 

Although the qualitative behavior of the AFM susceptibility is already correctly captured in the $N_{\text{FF}}=1$ calculation, there are slight quantitative differences. The inverse susceptibilities from the $N_{\text{FF}}=9$ calculations lie below those from calculations with $N_{\text{FF}}=1$ indicating effectively stronger correlations for the former calculation. Consistently, the deviation from linear behavior is stronger for  $N_{\text{FF}}=9$, for the temperatures shown.

\subsection{Self-energy at weak coupling ($U=2t$)}
\begin{figure}[tb]
\centering
\includegraphics{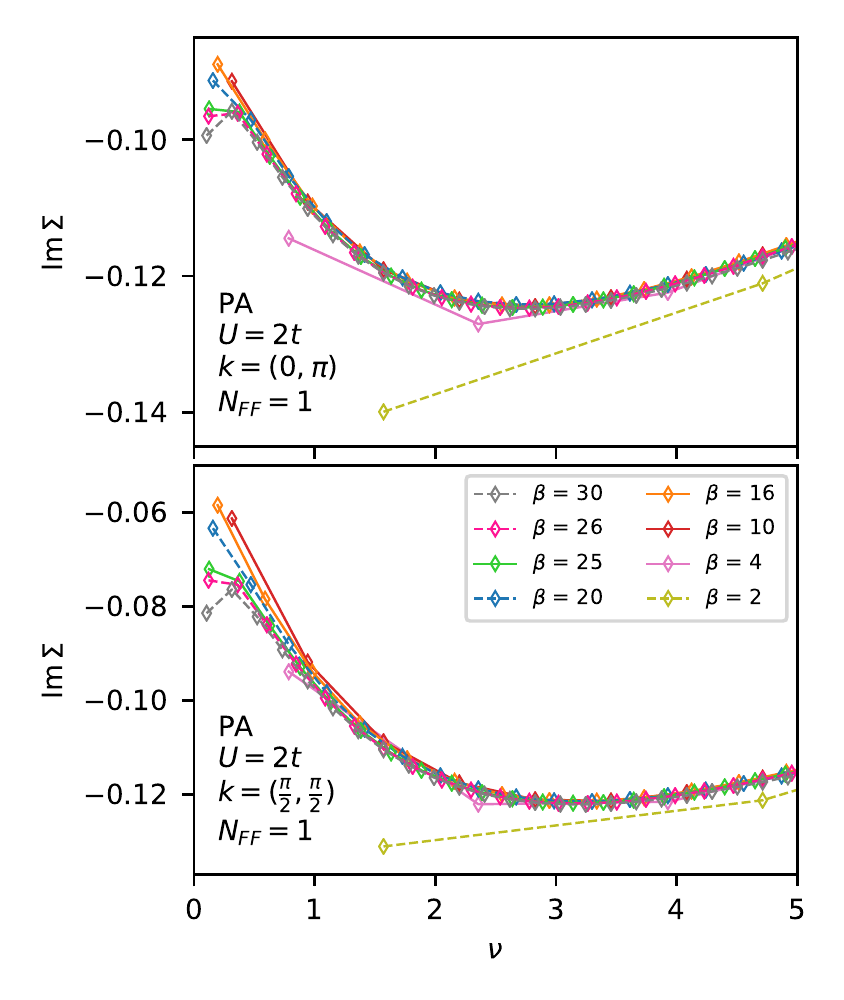}
\vspace{-5mm}
\caption{
Imaginary part of the self-energy $\text{Im} \, \Sigma(\vec k = (0, \pi), \nu_n)$ (top) and $\text{Im} \, \Sigma(\vec k = (\frac{\pi}{2}, \frac{\pi}{2}), \nu_n)$ (bottom) for two different momenta $\vec k$ as a function of Matsubara frequency $\nu_n$ for different temperatures in PA for the half-filled Hubbard model at $U=2t$, including a single basis function in the calculation ($N_{\text{FF}}=1$).
}
\label{fig:sig_U2_FF1}
\vspace{-0mm}
\end{figure}

\begin{figure}[tb]
\vspace{-0mm}
\setlength{\lineskip}{0pt}
\centering
\includegraphics{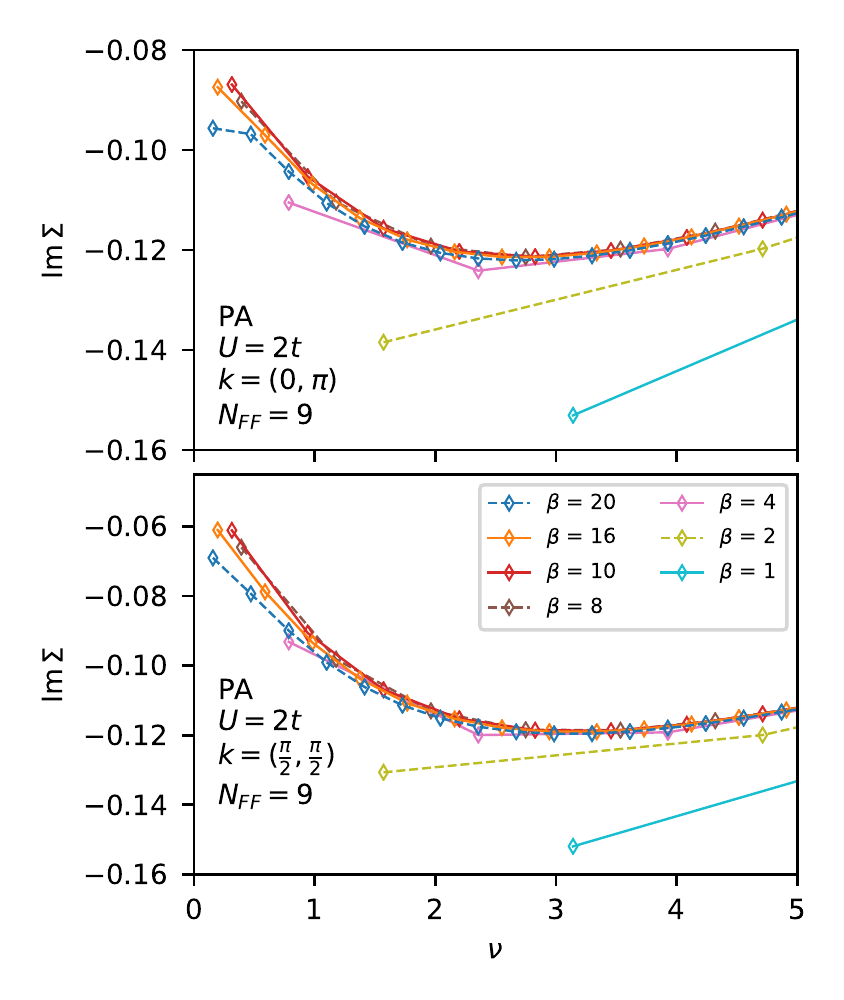}
\vspace{-5mm}
\caption{
Same as Fig.~\ref{fig:sig_U2_FF1} but for $N_{\text{FF}}=9$. In this case only for temperatures up to $\beta=20/t$ convergence in $\vec{q}$-grid could be achieved (with the $\beta=20/t$ point extrapolated to the infinite grid size). 
}
\label{fig:sig_U2_FF9}
\end{figure}

%
For a similar temperature range as for the susceptibilities in the previous Section we discuss the self-energy $\Sigma$ in the following. Again, we compare calculations with two different numbers of form factors $N_{\text{FF}}=1$ and $N_{\text{FF}}=9$. We show  $\text{Im} \, \Sigma(\vec k, \nu_n)$ for two points on the Fermi surface, $\vec k = (0, \pi)$ and $\vec k = (\frac{\pi}{2}, \frac{\pi}{2})$ at weak coupling ($U = 2t$) in Fig.~\ref{fig:sig_U2_FF1} for $N_{\text{FF}}=1$ and in Fig.~\ref{fig:sig_U2_FF9} for $N_{\text{FF}}=9$. Calculations have been performed in the PA. All results are converged in the number of Matsubara frequencies taken into account and in the number of discrete lattice momenta $\vec q$ down to $\beta = 16/t$.
For $\beta \ge 20/t$ we have extrapolated $\Sigma$ to infinite grid size.

For weak-coupling the qualitative behavior is also for the self-energy similar in the case of  $N_{\text{FF}}=1$ and $N_{\text{FF}}=9$; note that the  $N_{\text{FF}}=1$ case in  Fig.~\ref{fig:sig_U2_FF1} also includes much lower temperatures. 
 A simple first criterion for the physics of the solution, which is directly applicable on the Matsubara frequency axis, is the slope of the self-energy at low frequencies. A metallic Fermi liquid behavior implies a linear vanishing  of the imaginary part of self-energy   ${\rm Im}\Sigma(\nu)\stackrel{\nu\rightarrow 0}{\rightarrow}0$, in particular $|{\rm Im}\Sigma(\nu = \pi T)| < |{\rm Im} \Sigma(\nu = 3\pi T)|$ for the lowest two Matsubara frequencies. In the following discussion we drop the ``${\rm Im}$'' since for the half-filled particle-hole-symmetric Hubbard model considered ${\rm Re}\Sigma(\nu)=0$ on the Fermi surface {(with the Hartree part, equal to $U/2$, absorbed in the definition of the chemical potential)}.

At $\beta = 1/t$ and $\beta = 2/t$ we obtain $|\Sigma(\nu = \pi T)| > |\Sigma(\nu = 3 \pi T)|$ corresponding to a high temperature incoherent state.
Upon lowering the temperature a metallic behavior sets in,  indicated by $|\Sigma(\nu = \pi T)| < |\Sigma(\nu = 3\pi T)|$.
For the temperatures  shown here, one does not see that this crossover actually happens at different temperatures for $\vec k = (0, \pi)$ and $\vec k = (\frac{\pi}{2}, \frac{\pi}{2})$.

Upon further lowering the temperature we see $|\Sigma(\vec k = (0, \pi), \nu = \pi T)| > |\Sigma(\nu = 3 \pi T)|$ but at the same time, albeit only by a glimpse,  $|\Sigma(\vec k = (0, \pi), \nu = \pi T)| < |\Sigma(\nu = 3 \pi T)|$ at $\beta=26$ in  Fig.~\ref{fig:sig_U2_FF1} --- a regime known as pseudogap~\cite{Sadovskii1999}.
This behavior of the self-energy in the two dimensional Hubbard model, incoherent state $\rightarrow$ bad metal $\rightarrow$ pseudogap is consistently obtained by a wide range of computational methods to study correlated electron systems~\cite{Gull2013,SchaeferJMM,Thomas_paper,Kozik2018,Gukelberger2016} and therefore believed to be intrinsic to the Hubbard model. 
A further lowering of $T$  ($\beta=30$) indicates a completely gapped Fermi surface.

The onset of the pseudogap behavior is shifted to higher temperatures for $N_{\text{FF}}=9$  as compared to $N_{\text{FF}}=1$. For $N_{\text{FF}}=9$, we see first indications for pseudogap physics already at $\beta=20$ in  Fig.~\ref{fig:sig_U2_FF9} with  $|\Sigma(\nu = \pi T)| \approx |\Sigma(\nu = 3 \pi T)|$ at $\vec k = (0, \pi)$. Unfortunately, the larger numerical effort at  $N_{\text{FF}}=9$ form factors does not allow for lower temperatures properly converged  in the number of ${\vec q}$ points.
The narrow temperature window of the pseudogap for $U=2t$ is also observed in ladder D$\Gamma$A (cf. Fig.~5 of Ref.~\onlinecite{SchaeferJMM}), but the pseudogap onset is at a slightly higher temperature there ($T\approx t/18$)~\cite{SchaeferJMM}.   


%

Comparing $\Sigma$ from both $N_{\text{FF}}=1$ and $N_{\text{FF}}=9$ calculations at the lowest achieved temperature of $T = t/20$ we see that in the calculation with a single form-factor basis function one is further away from the pseudogap crossover. This fact is consistent with our previous finding in Sec.~\ref{sec:inverse_susceptibility} that the calculation with a single basis function leads to less correlated results. Quantitatively,  the number of form factors taken into account matters.

\subsection{Self-energy at intermediate-to-strong coupling ($U=4t$)}
\begin{figure}[tb]
\centering
\includegraphics{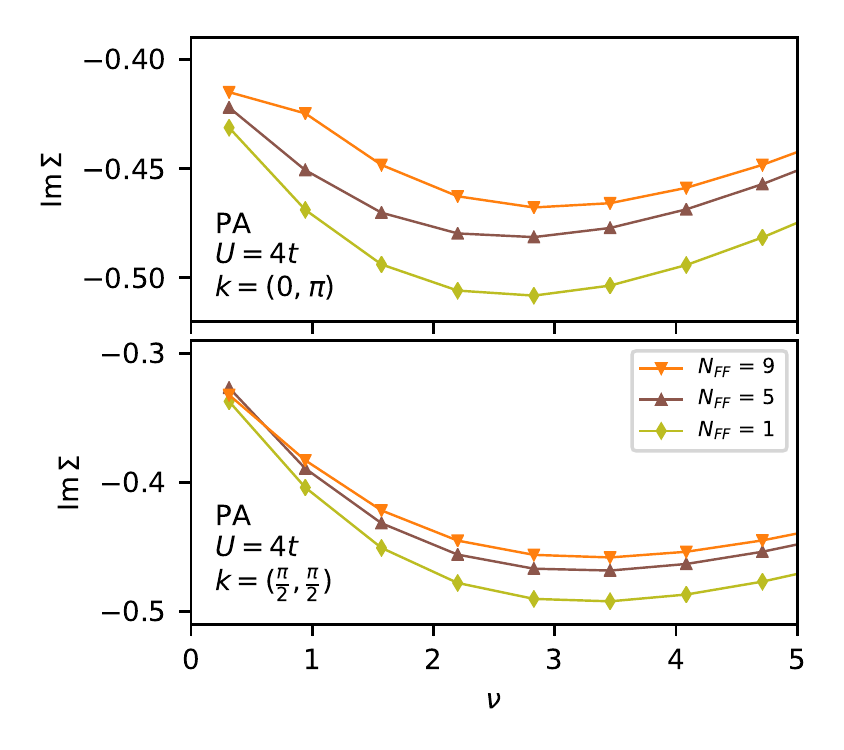}
\vspace{-5mm}
\caption{
$\text{Im} \, \Sigma(\nu_n)$ for $\vec k = (0, \pi)$ (top) and $\vec k = (\frac{\pi}{2}, \frac{\pi}{2})$ (bottom) in PA as function of Matsubara frequency $\nu_n$ at $U=4t$, $\beta = 10/t$ for different numbers of basis functions $N_{\text{FF}}$ considered in the calculation, 
$N_x \times N_y = 40 \times 40$ points in the Brillouin zone, and $N_{f+} = 40$ positive Matsubara frequencies.
}
\label{fig:sig_U4_PA}
\end{figure}
%
\begin{figure}[tb]
\centering
\includegraphics{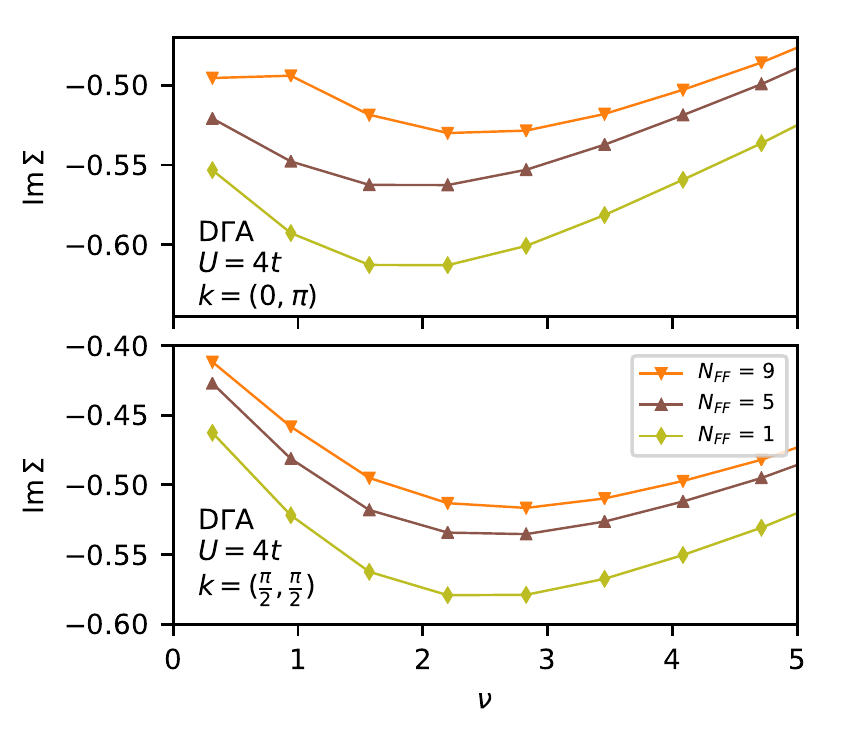}
\vspace{-5mm}
\caption{
Same as Fig.~\ref{fig:sig_U4_PA} but for D$\Gamma$A.
}
\label{fig:sig_U4_DGA}
\end{figure}

In the case of $U=4t$ we cannot perform a full analysis of the temperature behavior of the self-energy, since here for the same temperature range the correlation length is larger and the number of ${\vec q}$ points needed for convergence is higher than in the weak coupling case. 
We therefore focus in the following on the differences between calculations with different number of form factors at a fixed temperature of $T=t/10$.    

In Fig.~\ref{fig:sig_U4_PA} and Fig.~\ref{fig:sig_U4_DGA} we show results obtained with the PA and D$\Gamma$A, respectively. 
All calculations were performed taking $N_x \times N_y = 40 \times 40$ points in the BZ and $N_{f+} = 40$ positive fermionic Matsubara frequencies with which results are converged to  $2\%$ accuracy. We include results from calculations with $N_{\text{FF}}=1$,  $N_{\text{FF}}=5$ and  $N_{\text{FF}}=9$.
The results obtained with  D$\Gamma$A differ notably from those obtained with the PA. This is to be expected since setting $\Lambda = U$ in PA is valid only in the weak coupling regime, whereas the approximation for $\Lambda$ in D$\Gamma$A includes all local diagrams in $\Lambda$. The local dynamical correlations that become more important with increasing $U$ are also reflected in the dynamical structure of $\Lambda$ and the PA does not seem to be sufficient. Generally, the self-energies resulting from the D$\Gamma$A are larger in absolute values than those from PA and the effect of taking fewer form factors is also much stronger. The biggest difference is visible for $N_{\text{FF}}=9$, for which the self-energy shows already the pseudogap behavior in the case of the D$\Gamma$A calculation in Fig.~\ref{fig:sig_U4_DGA} but not in the case of the PA in Fig.~\ref{fig:sig_U4_PA}.

Consistent with the weak coupling results, the pseudogap transition is shifted to lower temperatures when performing calculations with a single basis function.
The calculation with $N_{\text{FF}}=5$ qualitatively resembles that with a single basis function and the results lie between the $N_{\text{FF}}=1$ and $N_{\text{FF}}=9$ ones.
Quantitatively, the dependence on the number of formfactors is quite sizable in  Figs.~\ref{fig:sig_U4_PA} and \ref{fig:sig_U4_DGA}, which calls for carefully analyzing the convergence with respect to the number of form factors.

\section{\label{sec:quantitative}Quantitative evaluation of the Truncated Unity approximation}
In this section we present the results of such a convergence study with respect to the number of basis functions $N_{\text{FF}}$ used in the form factor expansion. This way, we also evaluate the error of the TU basis truncation. 
The results from calculations with a truncated basis are compared to the original $\vec k$-space solution obtained with the {\it victory} code~\cite{victory}.
This $\vec k$-space solution is equivalent to performing calculations with the full basis which also holds numerically within machine precision. In order to make the calculations with the full basis set feasible, we chose an $8 \times 8$ momentum grid and a temperature of $T=t/5$. We have verified that the convergence in the number of form factors does not strongly depend on the size of the $\vec q$-grid making this study conclusive.

In the following we present the self-energy and static magnetic susceptibility calculated with different number of form factors $N_{\text{FF}}$ in the D$\Gamma$A for the same two values of $U$ as before, namely $2t$ and $4t$. The results of the PA for exactly the same parameters are included in the Appendix~\ref{App:E} for completeness. Our conclusions here only slightly depend on whether PA or D$\Gamma$A was used, and in the case of such dependence we explicitly mention it. For $U=2t$ the results of PA and D$\Gamma$A practically lie on top of each other.

\subsection{Weak coupling $U=2t$}
%
\begin{figure}[tb]
\centering
\includegraphics{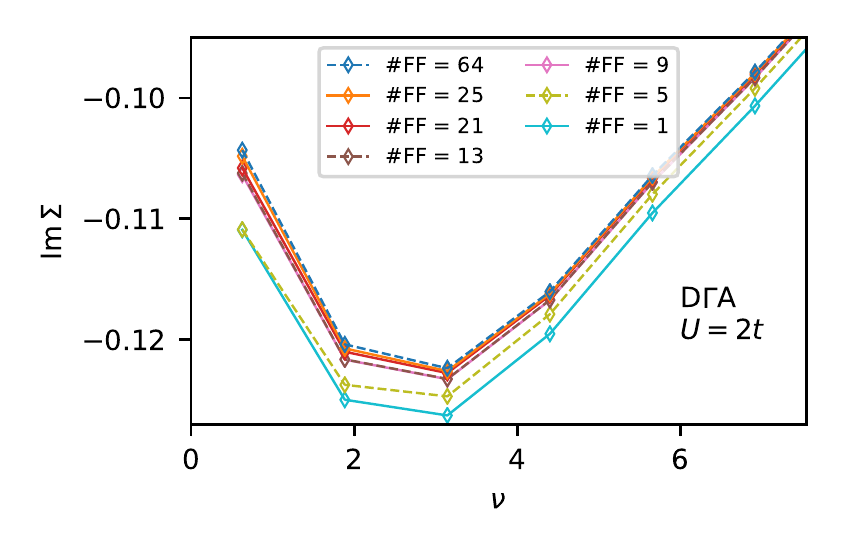}
\vspace{-5mm}
\caption{
$\text{Im} \, \Sigma(\vec k = (0, \pi), \nu_n)$ as function of Matsubara frequency $\nu_n$ for different numbers of basis functions $N_{\text{FF}}$ taken into account in D$\Gamma$A (PA results for the same parameters are presented in  Appendix \protect \ref{App:E}.
$U = 2t$, $\beta = 5/t$, $N_x \times N_y = 8 \times 8$, $N_{f+} = 16$.
\label{fig:sig_FF_U2_DGA}}
\end{figure}

\begin{figure}[t]
\centering
\begin{tikzpicture}
	\filldraw[black] (0,0) circle (1pt);
	\filldraw[black] (1,0) circle (1pt);
	\filldraw[black] (1,1) circle (1pt);
	\filldraw[black] (2,0) circle (1pt);
	\filldraw[black] (2,1) circle (1pt);
	\filldraw[black] (2,2) circle (1pt);
	\filldraw[black] (3,0) circle (1pt);
	\filldraw[black] (3,1) circle (1pt);
	\filldraw[black] (3,2) circle (1pt);
	\filldraw[black] (3,3) circle (1pt);
	\draw[thick, red, ->] (0.0, 0.0) -- (1.0, 0.0);
	\draw[thick, blue, ->] (0.0, 0.0) -- (1.0, 1.0);
	\draw[thick, green, ->] (0.0, -0.1) -- (2.0, -0.1);
	\draw[thick, yellow, ->] (0.0, 0.0) -- (2.0, 1.0);
	\draw[thick, magenta, ->] (-0.07, 0.07) -- (1.93, 2.07);
	\draw[thick, cyan, ->] (0.0, -0.2) -- (3.0, -0.2);
	\filldraw[red] (-0.5,3) circle (1pt) node[anchor=east] {$1^{\text{st}}$ $\hat{=} N_{\text{FF}} = 5$};
	\filldraw[blue] (-0.5,2.5) circle (1pt) node[anchor=east] {$2^{\text{nd}}$ $\hat{=} N_{\text{FF}} = 9$};
	\filldraw[green] (-0.5,2) circle (1pt) node[anchor=east] {$3^{\text{rd}}$ $\hat{=} N_{\text{FF}} = 13$};
	\filldraw[yellow] (-0.5,1.5) circle (1pt) node[anchor=east] {$4^{\text{th}}$ $\hat{=} N_{\text{FF}} = 21$};
	\filldraw[magenta] (-0.5,1) circle (1pt) node[anchor=east] {$5^{\text{th}}$ $\hat{=} N_{\text{FF}} = 25$};
	\filldraw[cyan] (-0.5,0.5) circle (1pt) node[anchor=east] {$6^{\text{th}}$ $\hat{=} N_{\text{FF}} = 29$};
\end{tikzpicture}
\vspace{2mm}
\vspace{-5mm}
\caption{Layers of neighbors on the square lattice. Only the irreducible part is shown.}
\label{fig:neighbours}
\end{figure}
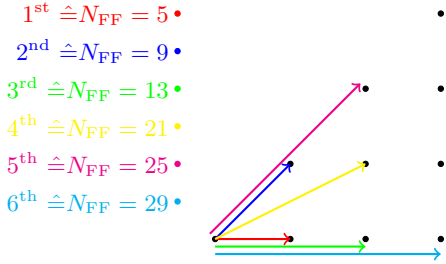

We begin by showing in Fig.~\ref{fig:sig_FF_U2_DGA} the self-energy $\Sigma$ at weak coupling $U = 2t$ as a function of Matsubara frequency for one point on the Fermi surface $\vec k = (0, \pi)$ and for different numbers of form factors $N_{\text{FF}}$. The chosen numbers between $1$ and $64$ correspond to different  shells of neighbors in the real lattice as indicated in Fig.~\ref{fig:neighbours},   $N_{\text{FF}}=64$ recovers the full basis set of the $8\times 8$ momentum grid. 

\begin{figure}[tb]
\centering
\includegraphics{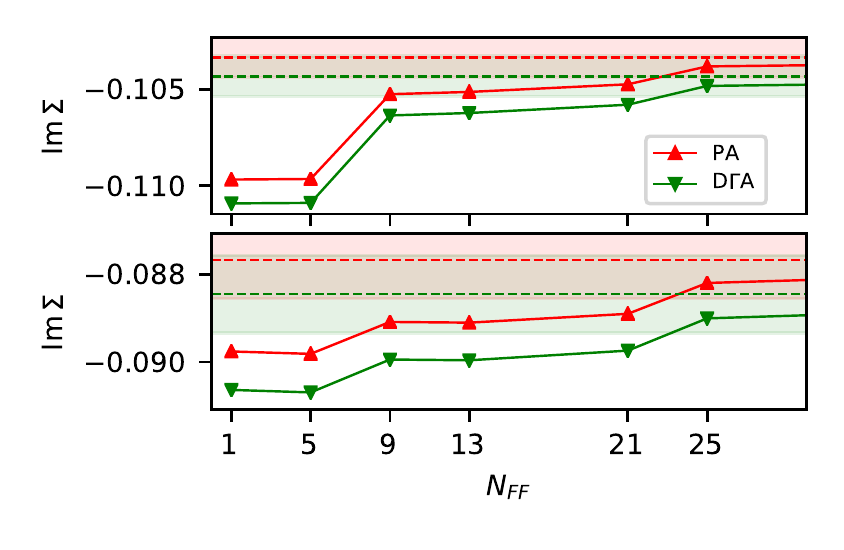}
\vspace{-5mm}
\caption{
$\text{Im}\,\Sigma(\vec k = (0, \pi), \nu_n = \pi T)$ (top) and $\text{Im}\,\Sigma(\vec k = (\frac{\pi}{2}, \frac{\pi}{2}), \nu_n = \pi T)$ (bottom) as function of the number of basis functions $N_{\text{FF}}$ used in the calculation. Both D$\Gamma$A and PA results are shown. Parameters as in Fig.~\ref{fig:sig_FF_U2_DGA}. The shaded area marks a $1\%$ difference to the full $\vec k$-space result indicated by the dashed line.
}
\label{fig:Sig_U2_Nlconv}
\end{figure}

While all results agree qualitatively, those for $N_{\text{FF}}=1$ and $N_{\text{FF}}=5$ differ quantitatively from the other curves.
With $N_{\text{FF}} = 9$ one accurately reproduces the full $\vec k$-space result. This is made more clear in Fig.~\ref{fig:Sig_U2_Nlconv} which shows $\Sigma$ for $\vec k = (0, \pi)$ and $\vec k = (\frac{\pi}{2}, \frac{\pi}{2})$ for the smallest fermionic Matsubara frequency $\nu = \pi T$.
The shaded area marks a $1\%$ difference to the full $\vec k$-space result indicated by the dashed line. In Fig.~\ref{fig:Sig_U2_Nlconv} also the PA data are shown confirming the agreement of PA and D$\Gamma$A for this value of $U$.
%
%
\begin{figure}[tb]
\centering
\includegraphics{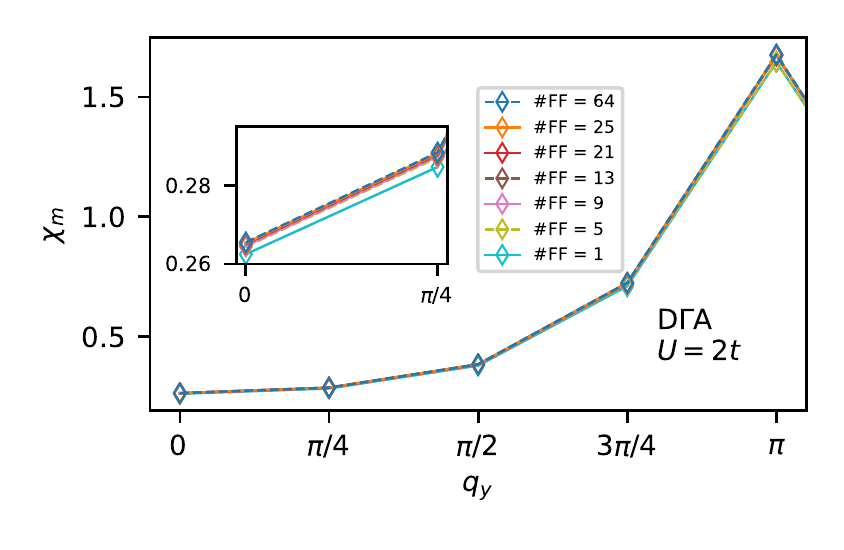}
\vspace{-5mm}
\caption{
Static magnetic susceptibility $\chi_m(\vec q = (\pi, q_y), \omega = 0)$ as function of lattice momentum transfer $q_y$ for different numbers of basis functions $N_{\text{FF}}$ taken into account in D$\Gamma$A. PA results are show in  Appendix ~\ref{App:E}. The parameters as in Fig.~\ref{fig:sig_FF_U2_DGA}.
}
\label{fig:chi_FF_U2_DGA}
\end{figure}

\begin{figure}[tb]
\centering
\includegraphics{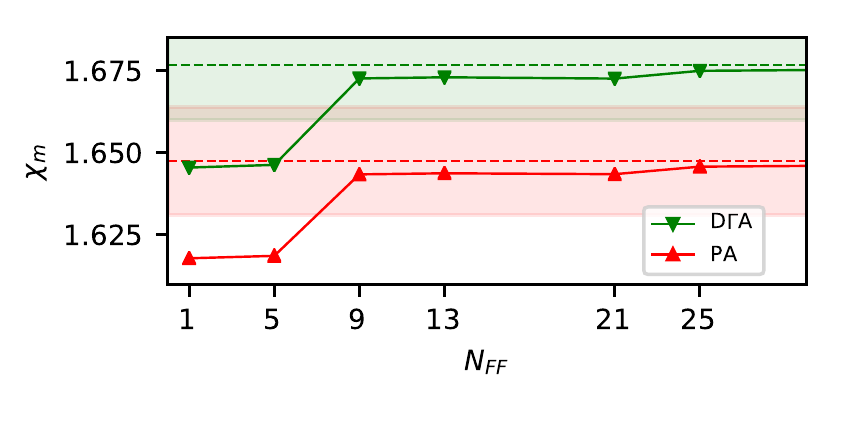}
\vspace{-5mm}
\caption{
Static AFM susceptibility as a function of the number of basis functions $N_{\text{FF}}$ used in the calculation.
Shaded area marks a $1\%$ difference to the $\vec k$-space value indicated by the dashed line. Both D$\Gamma$A and PA results are shown. The parameters as in Fig.~\ref{fig:sig_FF_U2_DGA}.
}
\label{fig:chi_U2_Nlconv}
\end{figure}

In Fig.~\ref{fig:chi_FF_U2_DGA} we further show the convergence in the number of basis functions of a 2-particle quantity namely the static magnetic susceptibility $\chi_m(\vec q = (\pi, q_y), \omega = 0)$  as a function of $q_y$ (with the AFM susceptibility $\chi_{\text{AFM}}$ being  at $q_y = \pi$). In this figure one cannot see any significant differences for calculations for different  numbers of basis functions.
For the point where the differences are the largest, i.e. $\chi_{\text{AFM}}$, we present the explicit dependence on the number of form factors  in Fig.~\ref{fig:chi_U2_Nlconv}. Again we add the PA results to show the small difference between the two methods in weak coupling. The shaded area in the plot indicates a $1\%$ difference to the original $\vec k$-space result.
With $N_{\text{FF}} = 1$ one already reproduced the $\vec k$-space result within $1-2\%$ accuracy and with $N_{\text{FF}} = 9$ the result can be considered converged in the number of form factors.

{\it False convergence.}
We find it important to  point out a peculiarity when checking convergence in the number of form factors.
When considering quantities for frequencies $\nu = \pi T$  or $\omega = 0$ as in Fig.~\ref{fig:Sig_U2_Nlconv}  and Fig.~\ref{fig:chi_U2_Nlconv} one sees that results with $N_{\text{FF}} = 1$ differ only insignificantly form those with $N_{\text{FF}} = 5$.
One could thus, in a less thorough analysis, be lead to believe to have had a converged result with only a single basis function if the test for convergence is comparison of the above quantities to a calculation with only one more basis function or neighbor in the real lattice. A similar effect has also been noted in Ref.~\onlinecite{Lichtenstein17}.

\subsection{Intermediate-to-strong coupling $U=4t$}

We continue to perform the same convergence analysis at $U = 4t$ for otherwise the same parameters as in the $U=2t$ case.
\begin{figure}[tb]
\centering
\includegraphics{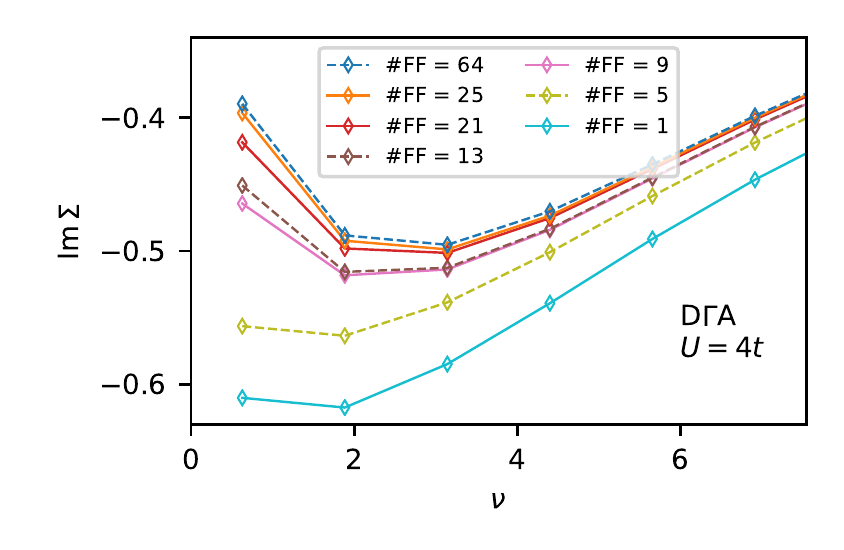}
\vspace{-5mm}
\caption{Same as in Fig.~\ref{fig:sig_FF_U2_DGA} but for $U=4t$. 
}
\label{fig:sig_FF_U4_DGA}\label{fig:Sig_FF_U4_DGA}
\end{figure}
In Fig.~\ref{fig:Sig_FF_U4_DGA} we show the self-energy $\Sigma$ for $\vec k = (0, \pi)$ as a function of imaginary frequency obtained in D$\Gamma$A with different number of form factors.
The PA results can be found again in Appendix~\ref{App:E}. For $U=4t$ the differences between PA and D$\Gamma$A are more pronounced, as already discussed in Sec.~IV.
Contrary to the weak coupling case, calculations with $N_{\text{FF}}=1$ and $N_{\text{FF}}=9$ differ significantly from the full $\vec k$-space solution.
The main qualitative difference is that the results with $N_{\text{FF}}=1$ and $N_{\text{FF}}=5$ form factors suggest an almost insulating or at least strongly incoherent behavior, whereas the results for the largest number of form factors indicate a Fermi liquid metal. Further, the maximum of $|\text{Im}\, \Sigma|$ shifts from $\nu = 5 \pi T$ to $\nu = 3 \pi T$.
\begin{figure}[tb]
\centering
\includegraphics{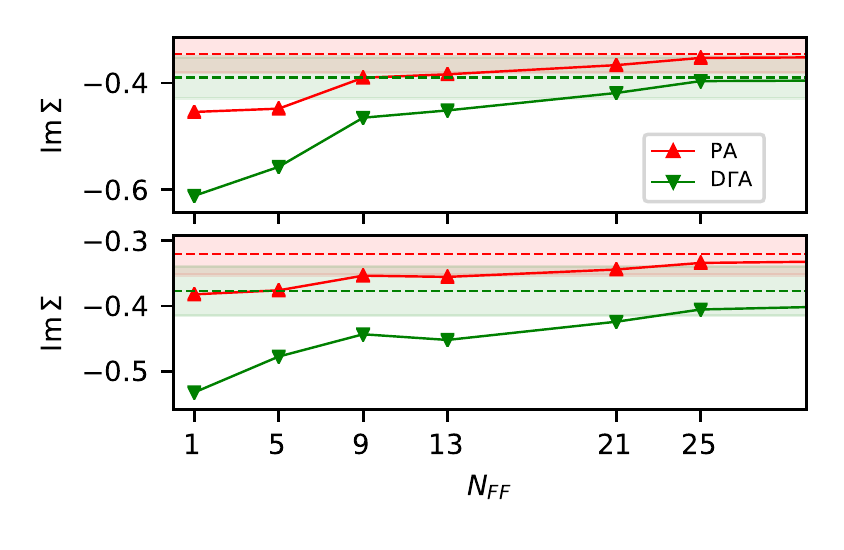}
\vspace{-5mm}
\caption{Same as in Fig.~\ref{fig:Sig_U2_Nlconv}, but for $U=4t$. The shaded area now marks a larger $10\%$ difference to the full $\vec k$-space result.
}
\label{fig:Sig_U4_Nlconv}
\end{figure}

Analogously to the weak coupling case we also show a direct dependence of the smallest frequency value of the self-energy on the number of form factors. In Fig.~\ref{fig:Sig_U4_Nlconv} both the results for D$\Gamma$A and PA are shown. The shaded area marks a $10\%$ difference to the full $\vec k$-space result. The results show that one needs more form factors in the intermediate-to-strong coupling regime than in the weak coupling regime to reach the same quantitative accuracy.
%
%
\begin{figure}[tb]
\centering
\includegraphics{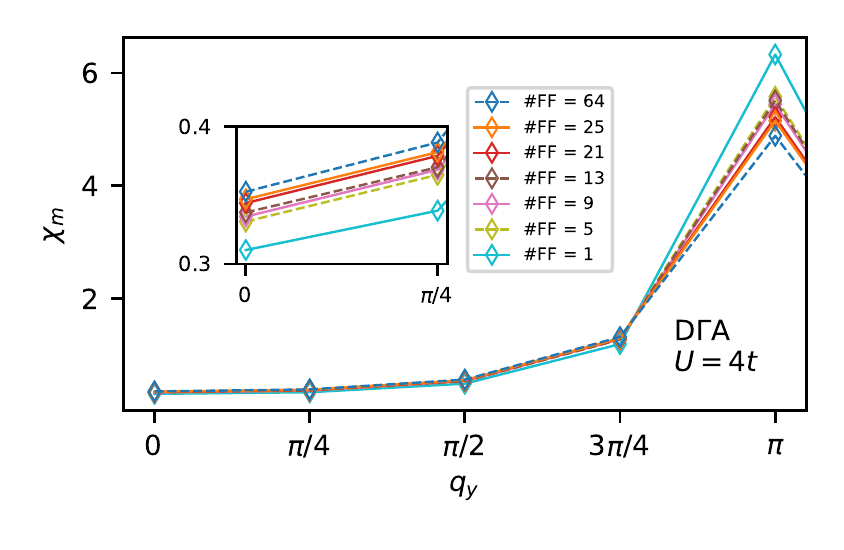}
\vspace{-5mm}
\caption{
Same as Fig.~\ref{fig:chi_FF_U2_DGA} but for $U=4t$.
}
\label{fig:chi_FF_U4_DGA}
\end{figure}

 We continue by considering the static magnetic susceptibility $\chi_m(\vec q = (\pi, q_y), \omega = 0)$ for different numbers of basis functions shown in Fig.~\ref{fig:chi_FF_U4_DGA}. Differences for the different calculations are now visible, being most pronounced at $\vec q = (\pi, \pi)$.
We notice that the AFM susceptibility decreases at $\vec q = (\pi, \pi)$ when performing calculations with more basis functions in contrast to the case of $U = 2t$ or other points in the BZ. This shows that the general trend of convergence is not predetermined. 

\begin{figure}[tb]
\centering
\includegraphics{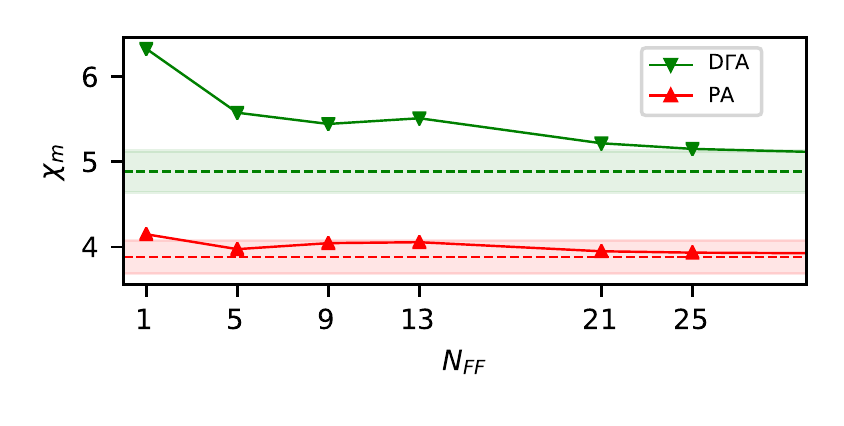}
\caption{
Same as Fig.~\ref{fig:chi_U2_Nlconv} but for $U=4t$.
Shaded area marks a $5\%$ difference to the $\vec k$-space value.
}
\label{fig:chi_U4_Nlconv}
\end{figure}

In Fig.~\ref{fig:chi_U4_Nlconv} we present again $\chi_{\text{AFM}}$ as function of number of form factors used in the calculation for PA and D$\Gamma$A. Here the two methods give quantitatively different values of $\chi_{\text{AFM}}$ and for the PA the convergence is faster than in the D$\Gamma$A case.
These results also show that the convergence in $N_{\text{FF}}$ does not need to be a monotonous function indicating that when performing convergence studies in form factors away from weak coupling special care is needed.



\section{Conclusion}
\label{Sec:conclusion}
We have implemented a truncated unity parquet solver (TUPS) which maps the fermionic momenta onto a form-factor basis with an arbitrary number of form-factors. The great advantage compared to previous parquet solvers \cite{Tam2013,Yang2009,Li2016,Kauch2019,Li2019} is that we arrive this way at a rather compact representation, where fewer form factors need to be taken into account than  momenta in previous parquet implementations. This saves considerable computing resources, in particular memory -- even to the extent that previous parquet codes were restricted by memory requirements, which is not the case for TUPS any more. TUPS, for the square lattice Hubbard model, is made available to the scientific community under GNU public license at GitHub~\cite{TUPS}. Input is  the fully irreducible vertex $\Lambda$ which might be taken as the bare $U$ as in the PA. Alternatively, one may take all local Feynman diagrams as in D$\Gamma$A.

With TUPS we were able to reach low enough temperatures to observe pseudogap physics. To the best of our knowledge, hitherto pseudogap physics has not been found in PA.
In fRG, indications of pseudogap physics have been seen in special cases~\cite{Rohe2005, Katanin2005a} but a systematic understanding is not yet available~\footnote{{For a recent truncated unity fRG (TU-fRG) work on this topic see~\cite{Hille2020a}}}.
We find that the pseudogap crossover temperature is lower when using only a single form factor. However, in the weak coupling regime, the qualitative description of the 2D Hubbard model with only one basis function is still possible.  In the intermediate to strong coupling regime, on the other hand, the quantitative differences between results obtained with only one form factor and the full basis are significant, particularly in the D$\Gamma$A calculations.

 An important further observation is that the inverse AF susceptibility in PA suscetibility bends away  from a linear mean-field behavior at low temperatures. This hints that the Mermin-Wagner\cite{Mermin1966} theorem is actually fulfilled in parquet-based approaches. While this has been conjectured before~\cite{Bickers1992}, our results provide, to the best of our knowledge,  the first numerical indication.

Our furthergoing systematic analysis confirms a quantitative dependence of the results on the number of form factors, which is more pronounced for a larger interaction $U$.  This is also a valuable observation for the fRG community, as  including a single form factor or very few form factors, as previously done, has to be taken with  a grain of salt. Calculations with only few form factors can serve for a computationally feasible qualitative analysis which treats all scattering channels on an equal footing.

\begin{acknowledgments}
We thank  S.~Andergassen, J.~Ehrlich, C.~Hille, F.~H\"orbinger, J.~Kaufmann, F.~Kugler, A.~Toschi, and C.~Watzenb\"ock  for valuable discussions. This work has been supported in part by  the Austrian Science Fund (FWF) through projects P\,30997 and P\,32044. CE and CH acknowledge support by Deutsche Forschungsgesellschaft through the project DFG-RTG 1995. Calculations have been done on the Vienna Scientific Cluster (VSC) and JURECA at Forschungszentrum J\"ulich.
\end{acknowledgments}

\appendix

\section{\label{app:su2}\label{sec:su2}SU(2) symmetry to simplify the parquet equations}
\label{App:A}

In the main paper, we did not include the spin indices since the form-factor expansion involves the momentum space only. Including the spin would require extra indices and equations,  only distracting from the main point. In the following we briefly indicate how the spin needs to be included.
 
One can show that for a SU(2) symmetric action for fermions,   obeying the Pauli-principle, the so-called crossing symmetries hold for the vertex functions.~\cite{Bickers04, Salmhofer00}
One can use it in order to restrict the calculation of vertex functions to a limited number of spin configurations since all other combinations can then be obtained by applying the crossing symmetry to the known quantities.
For this we introduce vertices for certain spin and momentum  combinations
\begin{equation}
\begin{aligned}
V_{d/m}^{k, k', q} := V_{\uparrow \uparrow \uparrow \uparrow}^{k, k'{+}q, k', k{+}q} +{/}- V_{\uparrow \downarrow \downarrow \uparrow}^{k, k'{+}q, k', k{+}q}\\
V_{t/s}^{k, k', s} := V_{\uparrow \downarrow \downarrow \uparrow}^{k, s{-}k, k', s{-}k'} +{/}- V_{\uparrow \downarrow \uparrow \downarrow}^{k, s{-}k, k', s{-}k'}
\end{aligned}
\label{eq:def_dmst}
\end{equation}
which we call \underline{d}ensity, \underline{m}agnetic, \underline{s}inglet and \underline{t}riplet.
All vertex quantities can be calculated from $\Lambda$ and the 2-particle reducible vertices $\Phi_{\text{ph}, d}$, $\Phi_{\text{ph}, m}$, $\Phi_{\text{pp}, s}$ and $\Phi_{\text{pp}, t}$ by means of the crossing symmetry and the BSE and PE.
In what follows we will abbreviate $\text{ph}, d \rightarrow d$, $\text{ph}, m \rightarrow m$, $\text{pp}, s \rightarrow s$ and $\text{pp}, t \rightarrow t$.
The corresponding PE for the irreducible vertices follow as~\cite{Bickers04,Rohringer2012,Li2016,Li2019}
\begin{widetext}
\begin{equation}
\begin{aligned}
\Gamma_{d}^{k_1, k_2}(q_1) &= \Lambda_{d}^{k_1, k_2}(q_1) - \frac{1}{2}\Phi_{d}^{k_1, k_1 + q_1}(k_2 - k_1) - \frac{3}{2}\Phi_{m}^{k_1, k_1+q_1}(k_2-k_1) + \frac{1}{2}\Phi_{s}^{k_1, k_2}(q_1+k_1+k_2) + \frac{3}{2}\Phi_{t}^{k_1, k_2}(q_1+k_1+k_2)\\
\Gamma_{m}^{k_1, k_2}(q_1) &= \Lambda_{m}^{k_1, k_2}(q_1) - \frac{1}{2}\Phi_{d}^{k_1, k_1 + q_1}(k_2 - k_1) + \frac{1}{2}\Phi_{m}^{k_1, k_1+q_1}(k_2-k_1) - \frac{1}{2}\Phi_{s}^{k_1, k_2}(q_1+k_1+k_2) + \frac{1}{2}\Phi_{t}^{k_1, k_2}(q_1+k_1+k_2)\\
\Gamma_{s}^{k_1, k_2}(q_1) &= \Lambda_{s}^{k_1, k_2}(q_1) + \frac{1}{2}\Phi_{d}^{k_1, k_2}(q_1 - k_1 - k_2) - \frac{3}{2}\Phi_{m}^{k_1, k_2}(q_1 - k_1 - k_2) + \frac{1}{2}\Phi_{d}^{k_1, q_1 - k_2}(k_2 - k_1) - \frac{3}{2}\Phi_{m}^{k_1, q_1 - k_2}(k_2 - k_1)\\
\Gamma_{t}^{k_1, k_2}(q_1) &= \Lambda_{t}^{k_1, k_2}(q_1) + \frac{1}{2}\Phi_{d}^{k_1, k_2}(q_1 - k_1 - k_2) + \frac{1}{2}\Phi_{m}^{k_1, k_2}(q_1 - k_1 - k_2) - \frac{1}{2}\Phi_{d}^{k_1, q_1 - k_2}(k_2 - k_1) - \frac{1}{2}\Phi_{m}^{k_1, q_1 - k_2}(k_2 - k_1) \; .
\end{aligned}
\label{eq:parquet_SU2}
\end{equation}
\end{widetext}

\section{\label{app:PE_opt}Further optimization of the parquet equation}
\label{App:B}
To better understand the extra approximations that come with Eq.~\eqref{eq:trafo_PE} (i.e., the occurrence of  non-channel-native momenta in the PE) but also to show how to optimally evaluate the equations we will perform another transformation of the 2-particle reducible vertex functions.
We thus define
\begin{equation}
\Phi_{r}^{(\vec k, \nu)(\vec k', \nu')(\vec R, \omega)} := \frac{1}{\sqrt{N}} \sum_{\vec q \in \text{1.BZ}}e^{i \vec q \cdot \vec R}%
\Phi_{r}^{(\vec k, \nu)(\vec k', \nu')(\vec q, \omega)} 
\label{eq:trafo_Phi}
\end{equation}
and conversely we have
\begin{equation}
\Phi_{r}^{(\vec k, \nu)(\vec k', \nu')(\vec q, \omega)} := \frac{1}{\sqrt{N}} \sum_{\vec q \in \text{1.BZ}}e^{-i \vec q \cdot \vec R}%
\Phi_{r}^{(\vec k, \nu)(\vec k', \nu')(\vec R, \omega)}. 
\label{eq:trafo_Phi_inv}
\end{equation}
Here $\vec R$ denotes a vector from the real-space lattice, we thus simply transform the bosonic momentum argument of the $\Phi_r$ into the real-space basis.
We stress the fact that the transformation in Eq.~\eqref{eq:trafo_Phi} does \emph{not} come with an extra approximation, since we use the whole real-space basis and \emph{not} a truncated basis.\\
We continue by inserting Eq.~\eqref{eq:trafo_Phi_inv} into Eq.~\eqref{eq:trafo_PE} which yields
\begin{widetext}
\begin{equation}
\begin{aligned}
\Gamma_{d}&^{(\ell_1, \nu_1), (\ell_2, \omega_2 + \nu_1), (\vec q_1, \omega_1) }\\
&\overset{1.}{\leftarrow}
\sum_{\ell_3 \ell_4} \sum_{\vec q_2} 
\left[ -\frac{1}{2} \Phi_{d} - \frac{3}{2} \Phi_{m} \right]^{(\ell_3, \nu_1), (\ell_4, \omega_1 + \nu_1), (\vec q_1, \omega_1)}
\sum_{\vec k_1} f^{\ell_1, \vec k_1} f^{\ell_2, \vec k_1 + \vec q_2}
f^{\ell_3, \vec k_1} f^{\ell_4, \vec k_1 + \vec q_1}\\
&\overset{2.}{=}
\sum_{\ell_3 \ell_4} \sum_{\vec q_2} \sum_{\vec R_2}
\left[ -\frac{1}{2} \Phi_{d} - \frac{3}{2} \Phi_{m} \right]^{(\ell_3, \nu_1), (\ell_4, \omega_1 + \nu_1), (\vec R_2, \omega_2)}
\frac{1}{\sqrt{N}} e^{i \vec q_2 \cdot \vec R} \times \\
& \times \sum_{\vec k_1} f^{\ell_1, \vec k_1} f^{\ell_2, \vec k_1 + \vec q_2}
f^{\ell_3, \vec k_1} f^{\ell_4, \vec k_1 + \vec q_2}\\
&\overset{3.}{=}
\sum_{\ell_3 \ell_4} \sum_{\vec R_2}
\left[ -\frac{1}{2} \Phi_{d} - \frac{3}{2} \Phi_{m} \right]^{(\ell_3, \nu_1), (\ell_4, \omega_1 + \nu_1), (\vec R_2, \omega_2)} \times \\
& \times \frac{1}{\sqrt{N}} 
\sum_{\vec k_1} f^{\ell_1, \vec k_1} f^{\ell_3, \vec k_1} f^{\ell_4, \vec k_1 + \vec q_1} e^{ - i \vec k_1 \cdot \vec R}
\sum_{\vec q_2} f^{\ell_2, \vec k_1 + \vec q_2} e^{i (\vec k_1 + \vec q_2) \cdot \vec R}\\
&\overset{4.}{=}
\sum_{\ell_3 \ell_4} \sum_{\vec R_2}
\left[ -\frac{1}{2} \Phi_{d} - \frac{3}{2} \Phi_{m} \right]^{(\ell_3, \nu_1), (\ell_4, \omega_1 + \nu_1), (\vec R_2, \omega_2)} \times \\
& \times \frac{1}{\sqrt{N}} 
\sum_{\vec k_1} f^{\ell_1, \vec k_1} f^{\ell_3, \vec k_1} f^{\ell_4, \vec k_1 + \vec q_1} e^{ - i \vec k_1 \cdot \vec R_2}
\sum_{\vec k} f^{\ell_2, \vec k} e^{i \vec k \cdot \vec R_2}\\
\end{aligned}
\label{eq:trafo_PE_optimized}
\end{equation}
\end{widetext}
At step 1.) we have inserted the result from Eq.~\eqref{eq:trafo_PE}, at 2.) we inserted the transformation Eq.~\eqref{eq:trafo_Phi_inv}, at 3.) we reordered terms and in 4.) we substituted the occurring sums.
In order to further interpret this result consider the term 
\begin{equation}
\sum_{\vec k \in \text{1.BZ}} f^{\ell, \vec k} e^{i \vec q \cdot \vec R}.
\label{eq:vanishing_term}
\end{equation}
As stated in section Sec.~\ref{sec:tu_parquet} $f$ is given by a linear combination of complex exponential functions of the form $e^{i \vec r \cdot \vec k}$ where $\vec k$ is a vector from the 1.BZ and $\vec r$ a real space lattice vector. 
Each lattice vector $\vec r$ has a specific distance to the origin $d_{\ell}$ for any given $\ell$.
It, therefore, follows
\begin{equation}
\sum_{\vec k \in \text{1.BZ}} f^{\ell, \vec k} e^{i \vec q \cdot \vec R} = 0 ; \hspace{2mm} \text{for} \, |\vec R| > d_{\ell}.
\label{eq:vanishing_term_shown}
\end{equation}
Performing calculations in a reduced basis set means setting all basis function with index $\ell$ for which $d_{\ell}>d_{\ell_{\text{max}}}$ holds to zero.
We therefore see, that the sum $\sum_{\vec R_2}$ in Eq.~\eqref{eq:trafo_PE_optimized} only needs to be computed over the reduced part of the real lattice for which $|\vec R_2| < d_{\ell_{\text{max}}}$ holds.\\
The here shown arguments are generic for all contributions to the parquet equations and a similar truncation of the real-space lattice sum occurs for the other contributions in the same way.\\
We note that the last line in Eq.~\eqref{eq:trafo_PE_optimized} includes a truncated sum over the real lattice including as many terms as basis functions in the form-factor expansion.
On the other hand, the first line of Eq.~\eqref{eq:trafo_PE_optimized} includes a sum over the whole BZ.
Thus our extra transformation yields a PE which scales linearly in the number of discrete lattice momenta taken into account compared to the previous form which scaled quadratically. 
The above derivation also illustrates the extra approximation that comes with the necessity to turn around arguments in the PE.
The above shown example is a channel cross-insertion in particular the insertion of 2-particle reducible vertices from the $\ov{\text{ph}}$-channel into the ph-channel. For the last, bosonic argument, i.e.,  $\vec R_2$ in Eq.~(\ref{eq:trafo_PE_optimized}), can be very non-local as it corresponds to $\vec q$. However by transforming to  ${\ell}_1$ and ${\ell}_2$ with  form-factors such as  $f^{\ell_2, \vec k_1}$ one approximates this bosonic argument  $\vec R_2$ in a similar way as otherwise the fermionic arguments.
\section{\label{app:tu_sde}Different versions of the TU Schwinger-Dyson equation} 
In this Appendix we point to a numerical subtlety when performing computations in a reduced basis within the TU method.
This study will also further illustrate the approximations discussed in Sec.~\ref{sec:derivation_tu_parquet}.
In Sec.~\ref{sec:derivation_tu_sde} we introduced the TU version of the SDE as being split up in its different contributions from the fully 2P-irreducible vertex $\Lambda$ and the 2-particle reducible vertices $\Phi_{r}$ where $r = \text{ph}, \ov{\text{ph}}, \text{pp}$.
Alternatively, one can evaluate the SDE with the full $F$ as for example {(we omit the constant Hartree contribution in the following)}
\begin{equation}
\begin{aligned}
&\underline \Sigma = 
 - \frac{1}{2} \sum_{\vec q_1, \omega_1} \left[ \left[\mat{ F_d} -\mat{ F_m} \right]^{(\vec q_1, \omega_1)} \otimes \vec \chi_{0,\text{ph}}^{(\vec q_1, \omega_1)} \right] \cdot \underline G_{\text{ph}}^{(\vec q_1, \omega_1)}.
\end{aligned}
\label{eq:SDE_ph}
\end{equation}
The equivalence of Eq.~\eqref{eq:SDE_ph} and Eq.~\eqref{eq:SDE_split_up} is ensured by the PE Eq.~\eqref{eq:parquet_no_args}.
As discussed in App.~\ref{app:PE_opt} this equivalence does not strictly hold in the case of a truncated basis in a TU calculation due to the approximated channel coupling.
To illustrate this point we implemented a sup-optimal version of the SDE as
\begin{equation}
\begin{aligned}
&\underline \Sigma = \\
& - \frac{1}{6} \sum_{\vec q_1, \omega_1} \left[\left[ \mat{ F_d} -  \mat{F_m} \right]^{(\vec q_1, \omega_1)} \otimes \underline \chi_{0,\text{ph}}^{(\vec q_1, \omega_1)} \right] \cdot \underline G_{\text{ph}}^{(\vec q_1, \omega_1)}\\
& + \frac{1}{3} \sum_{\vec q_1, \omega_1} \left[ \mat{F}_m^{(\vec q_1, \omega_1)} \otimes \underline \chi_{0, \text{ph}}^{(\vec q_1, \omega_1)} \right] \cdot \underline G_{\text{ph}}^{(\vec q_1, \omega_1)}\\
& - \frac{1}{6} \sum_{\vec q_1, \omega_1} \left[ \left[ \mat{ F_s} -\mat{ F_t} \right]^{(\vec q_1, \omega_1)} \otimes \underline \chi_{0, \text{pp}}^{(\vec q_1, \omega_1)} \right] \cdot \underline G_{\text{pp}}^{(\vec q_1, \omega_1)}.
\end{aligned}
\label{eq:SDE_from_F}
\end{equation}
Here we calculate $\Sigma$ from the full $F$ but have mixed versions of the SDE written in all three channel-parametrizations in order to not be biased towards a specific channel.
We emphasize that the evaluation of Eq.~\eqref{eq:SDE_from_F} is \emph{not} more expensive than the evaluation of Eq.~\eqref{eq:SDE_split_up}.
\begin{figure}[tb]
\centering
\includegraphics{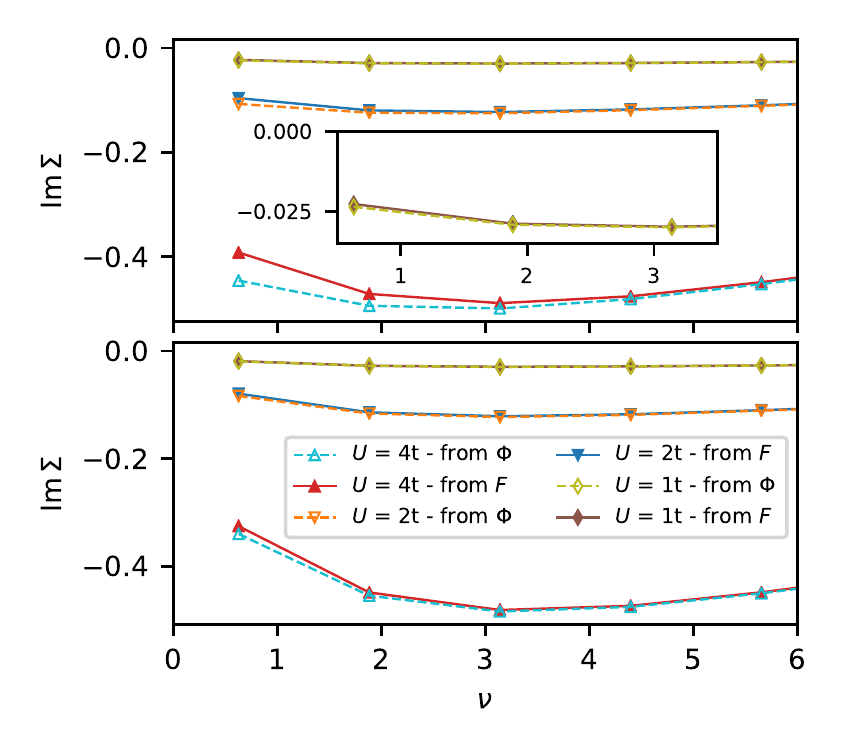}
\vspace{-5mm}
\caption{
$\Sigma(\nu)$ for $\vec k = (0, \pi)$ (top) and $\vec k = (\frac{\pi}{2}, \frac{\pi}{2})$ for different values of the Hubbard interaction $U$ at$\beta = 5/t$. 
Once $\Sigma$ was calculated with Eq.~\eqref{eq:SDE_transformed} from Sec.~\ref{sec:derivation_tu_sde} and once from Eq.~\eqref{eq:SDE_from_F}.
}
\label{fig:Sig_from_F}
\end{figure}
In Fig.~\ref{fig:Sig_from_F} we show $\Sigma$ for two points on the Fermi surface, $\vec k = (0, \pi)$ and $\vec k = (\frac{\pi}{2}, \frac{\pi}{2})$ as function of Matsubara frequency calculated with a single basis function $N_{\text{FF}} = 1$.
One sees that at $U=1t$ differences between the two calculations are negligible.
At $U=2t$ they become visible and more pronounced at $U=4t$. 
Qualitatively one notices, that computing $\Sigma$ from the full $F$ decreased the difference of the self-energy at $\vec k = (0, \pi)$ and $\vec k = (\frac{\pi}{2}, \frac{\pi}{2})$, i.e., the pseudogap phenomenon when compared to $\Sigma$ computed from the different $\Phi$.\\
\begin{figure}[tb]
\centering
\includegraphics{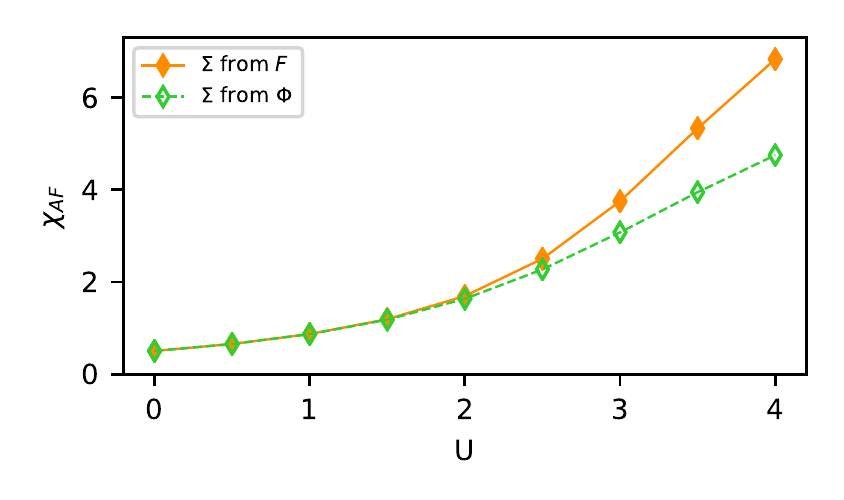}
\vspace{-5mm}
\caption{
Antiferromagnetic susceptibility $\chi_{\text{AF}}$ vs.\ Hubbard interaction $U$ at $\beta = 5/t$.
In one computation $\Sigma$ was calculated from $\Phi$ with Eq.~\eqref{eq:SDE_transformed} from Sec.~\ref{sec:derivation_tu_sde} and once from $F$ with Eq.~\eqref{eq:SDE_from_F}.
}
\label{fig:chiAF_Sig_from_F}
\end{figure}
We also show the effect of the two different implementations of the SDE on the AF susceptibility in Fig.~\ref{fig:chiAF_Sig_from_F} for the same calculation as above.
One sees that towards stronger interactions calculations from the sub-optimal implementation of the SDE yield increased susceptibilities.

Together with the self-energies this gives a consistent picture of the effects of using a sub-optimal SDE, where not every contribution is in its channel native momentum.
The further approximated channel coupling in a calculation with a single basis function leads to the AFM fluctuations in part only contributing locally to the self-energy in the SDE. 
Therefore, the pseudogap phenomenon is decreased as well as one obtains generally smaller values for $\Sigma$. 
This decreased $\Sigma$ in turn does not screen the vertex adequately yielding increased AFM susceptibilities.
When increasing the number of basis functions in the calculations we expect these differences to shrink. 

\section{Crossing symmetry \label{App:D}}

{In the following we will show how the truncated unity approximation influenes the crossing symmetry. We will generate a $k$-space vertex from a previously calculated $\ell$-space vertex and apply the crossing symmetry.
Let us define a doubly transformed vertex $V'$:
\begin{equation}
\begin{aligned}
V'^{\;k_1, k_3, q} &= \sum_{\ell_1, \ell_3} f^{*\; \ell_1, \vec k_1} \tilde{V}^{l_1, l_3, q} f^{\ell_3, \vec k_3 } \\
= &\sum_{\ell_1, \ell_3} f^{ *\;\ell_1, \vec k_1 } \left[ \sum_{\vec k'_1, \vec k'_3} f^{ \ell_1, \vec k'_1} V^{k'_1, k'_3, q} f^{*\;\ell_3, \vec k'_3}  \right] f^{ \ell_3, \vec k_3} 
\end{aligned}
\label{eq:def_V_dash}
\end{equation}
where $l = (\ell, \nu)$ and $k = (\vec k, \nu )$ are multiindices comprising frequency and momentum dependence and $q$ is the total momentum/frequency $q = k_1 + k_2$.
Note that for a truncated basis $V' \neq V$.
The crossing symmetry applied to the transformed vertex $V'$ would read
\begin{equation}
V'^{\;k_1, k_3, q} = - V'^{\;k_1, q - k_3, q}
\label{eq:crossing_transformed}
\end{equation}
We can check if this holds true, by taking the RHS of Eq.~\eqref{eq:crossing_transformed} and performing the double transformation from Eq.~\eqref{eq:def_V_dash}. Then we apply the known crossing symmetry to the original vertex $V$ and compare the result to $V'$ from equation Eq.~\eqref{eq:def_V_dash}.
\begin{widetext}
\begin{equation}
\begin{aligned}
- V'^{k_1, q - k_3, q} = &- \sum_{\ell_1, \ell_3} f^{*\; \ell_1, \vec k_1} \tilde{V}^{l_1, l_3, q} f^{\ell_3, \vec q -\vec k_3 } 
= - \sum_{\ell_1, \ell_3} f^{ *\;\ell_1, \vec k_1 } \left[ \sum_{\vec k'_1, \vec k'_3} f^{ \ell_1, \vec k'_1} V^{k'_1, k'_3, q} f^{*\;\ell_3, \vec k'_3}  \right] f^{ \ell_3, \vec q - \vec k_3} \\
\overset{\text{Cross. sym.}}{=} & - \sum_{\ell_1, \ell_3} f^{ *\;\ell_1, \vec k_1 } \left[ \sum_{\vec k'_1, \vec k'_3} f^{ \ell_1, \vec k'_1}\left (- V^{k'_1, q -k'_3, q}\right ) f^{*\;\ell_3, \vec k'_3}  \right] f^{ \ell_3, \vec q - \vec k_3} \\
\overset{\text{Subst.}}{=} &\sum_{\ell_1, \ell_3} f^{ *\;\ell_1, \vec k_1 } \left[ \sum_{\vec k'_1, \vec k''_3} f^{ \ell_1, \vec k'_1} V^{k'_1, k''_3, q} f^{*\;\ell_3, \vec q - \vec k''_3}  \right] f^{ \ell_3, \vec q - \vec k_3} \, ,
\end{aligned}
\label{eq:crossing_projection}
\end{equation}
\end{widetext}
where in the last part we substituted  $\vec k''_3 =\vec q - \vec k'_3$.
Comparing Eq.~\eqref{eq:def_V_dash} with the last line of Eq.~\eqref{eq:crossing_projection} one first notices that in the case of the full basis the crossing symmetry is again fulfilled. On the other hand, for a reduced basis, the equations differ. The reason is that for the transformation one picks out two fermionic arguments for the transformation, $k_1$ and $k_3$ in our case (cf. Eq.~\eqref{eq:vertex_fermionic_notation}), and transforms them. When applying the crossing symmetry, one exchanges two arguments ($k_3$ and $k_4$ in the way we formulated it now) thus the result looks as if one had picked $k_1$ and $k_4$ for the projection. In the truncated basis the crossing symmetry is thus violated.}

\section{\label{app:pa_results}Form-factor convergence in the PA} \label{App:E}
%
\begin{figure}[tb]
\centering
\includegraphics{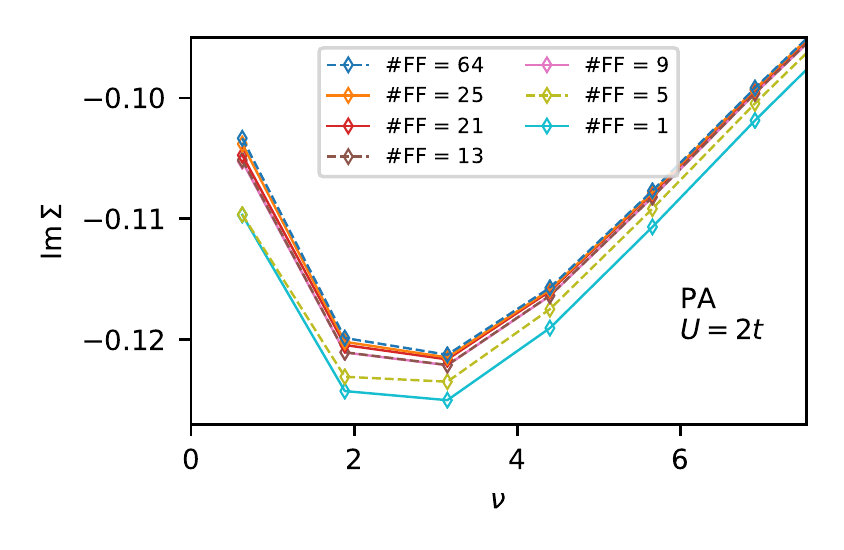}
\vspace{-5mm}
\caption{
$\text{Im} \, \Sigma(\vec k = (0, \pi), \nu_n)$ as function of Matsubara frequency $\nu_n$ for different numbers of basis functions $N_{\text{FF}}$ taken into account in PA. Same as Fig.~\ref{fig:sig_FF_U2_DGA} but using the PA instead of D$\Gamma$A. 
}
\label{fig:sig_FF_U2_PA}
\end{figure}

%
\begin{figure}[tb]
\centering
\includegraphics{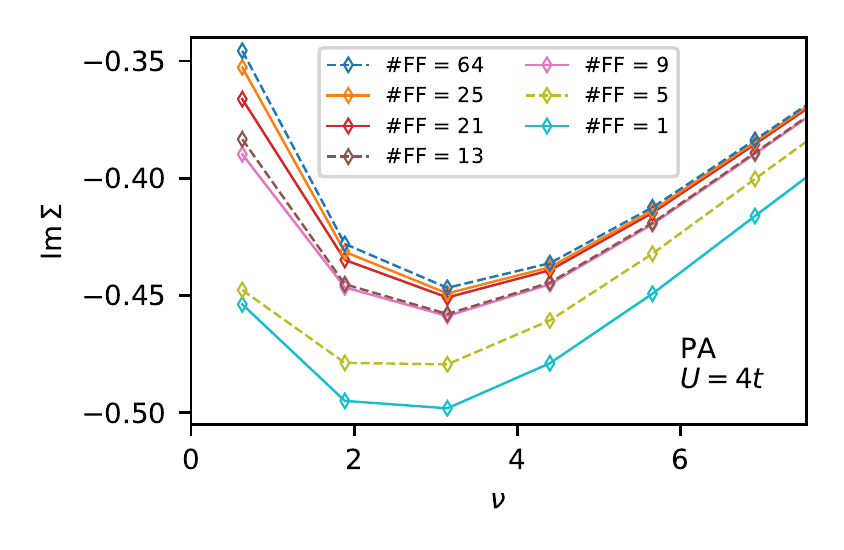}
\vspace{-5mm}
\caption{
$\text{Im} \, \Sigma(\vec k = (0, \pi), \nu_n)$ as function of Matsubara frequency $\nu_n$ for different numbers of basis functions $N_{\text{FF}}$ taken into account in PA.  Same as Fig.~\ref{fig:sig_FF_U4_DGA} but using the PA instead of D$\Gamma$A. 
}
\label{fig:sig_FF_U4_PA}
\end{figure}

%
\begin{figure}
\centering
\includegraphics{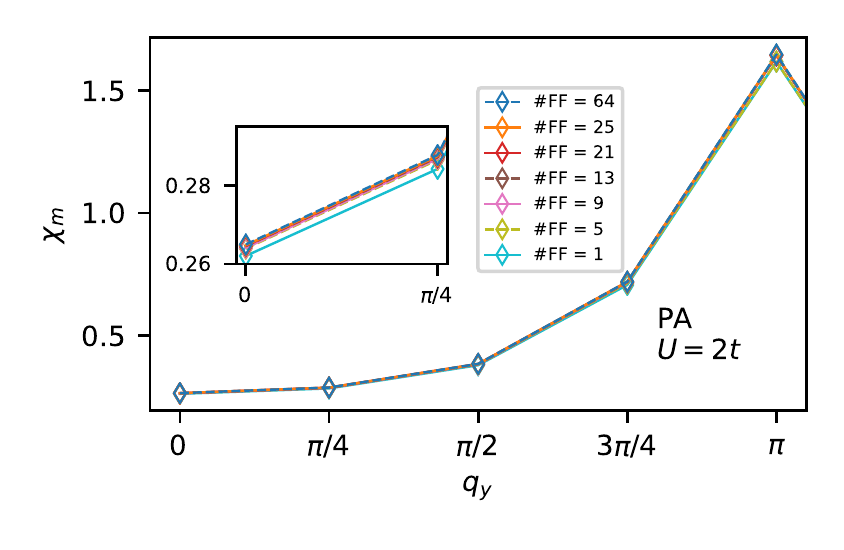}
\vspace{-5mm}
\caption{
$\chi_m(\vec q = (\pi, q_y), \omega = 0)$ as function of lattice momentum transfer $q_y$ for different numbers of basis functions $N_{\text{FF}}$ taken into account.  Same as Fig.~\ref{fig:chi_FF_U2_DGA} but using the PA instead of D$\Gamma$A. in PA.
}
\label{fig:chi_FF_U2_PA}
\end{figure}
%

\begin{figure}
\centering
\includegraphics{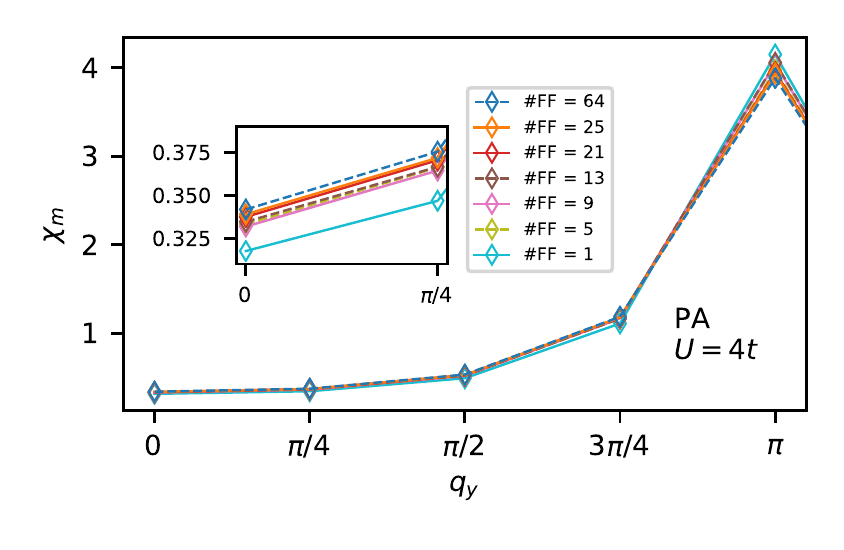}
\vspace{-5mm}
\caption{
$\chi_m(\vec q = (\pi, q_y), \omega = 0)$ as function of lattice momentum transfer $q_y$ for different numbers of basis functions $N_{\text{FF}}$ taken into account. Same as Fig.~\ref{fig:chi_FF_U4_DGA} but using the PA instead of D$\Gamma$A.
}
\label{fig:chi_FF_U4_PA}
\end{figure}

Finally we show in Figs.\ \ref{fig:sig_FF_U2_PA}, \ref{fig:sig_FF_U4_PA}, \ref{fig:chi_FF_U2_PA} and \ref{fig:chi_FF_U4_PA} the same results as in  Figs.\ \ref{fig:sig_FF_U2_DGA}, \ref{fig:sig_FF_U4_DGA}, \ref{fig:chi_FF_U2_DGA} and \ref{fig:chi_FF_U4_DGA} of the main text but now using the PA instead of the D$\Gamma$A.
While for weak coupling ($U=2t$) the PA and  D$\Gamma$A show very similar results,  there are quite substantial differences for intermediate-to-strong-coupling ($U=4t$). In particular the  D$\Gamma$A solution shows a somewhat larger antiferromagnetic susceptibility and consequently a more insulating self-energy.
Most notable and also relevant to the analysis of our paper, is the much stronger dependence on the number of form-factors in   D$\Gamma$A.


%
\end{document}